	\documentclass[draftclsnofoot, onecolumn, 12pt]{IEEEtran}   

	\usepackage{amsmath,amssymb,bm}
	\usepackage{subfigure,graphicx}

	\usepackage{tikz}
	\usetikzlibrary{automata,snakes} 

	\usepackage{citesort, url} 

	\newtheorem{theorem}{Theorem}
	
	\newtheorem{lemma}{Lemma}
	\newtheorem{proof}{Proof}
	\newtheorem{remark}{Remark}
	
	\allowdisplaybreaks
	\IEEEoverridecommandlockouts

\title{Paranoid Secondary: Waterfilling in a Cognitive Interference Channel with Partial Knowledge}
\author{
	\IEEEauthorblockN{Debashis Dash and Ashutosh Sabharwal}
	\thanks{The authors are with the 
		Department of Electrical and Computer Engineering, 
		Rice University, Houston, TX 77005, USA. 
		(email: \{ddash, ashu\}@rice.edu). 
		This work was partially supported by NSF Awards CCF-0635331 and CNS-0325971.
	}
}

\begin{document}
\maketitle

\begin{abstract}
We study a two-user cognitive channel, where the primary flow is sporadic, cannot be re-designed and operating below its link capacity. To study the impact of primary traffic uncertainty, we propose a \emph{block activity model} that captures the random on-off periods of primary's transmissions. Each block in the model can be split into parallel Gaussian-mixture channels, such that each channel resembles a multiple user channel (MAC) from the point of view of the secondary user. The secondary senses the current state of the primary at the start of each block. We show that the optimal power transmitted depends on the sensed state and the optimal power profile is \emph{paranoid}, i.e.\ either growing or decaying in power as a function of time. We show that such a scheme achieves capacity when there is no noise in the sensing. The optimal transmission for the secondary performs rate splitting and follows a \emph{layered water-filling} power allocation for each parallel channel to achieve capacity.  The secondary rate approaches a genie-aided scheme for large block-lengths. Additionally, if the fraction of time primary uses the channel tends to one, the paranoid scheme and the genie-aided upper bound get arbitrarily close to a no-sensing scheme.
\end{abstract}

\begin{IEEEkeywords} 
	Cognitive radio, spectrum sensing, interference channel, Gaussian mixture channel, capacity, side information, rate splitting, water-filling.
\end{IEEEkeywords}


\section{Introduction \label{sec:intro}}
Cognitive wireless is a novel approach to deploy new wireless services in the presence of legacy devices with priority access to the channel. The aim of a cognitive framework is to communicate as  an underlay to an underutilized primary network without degrading the primary communication beyond a predetermined threshold. 
The key constraint for our formulation is that the primary transceiver is legacy but fixed i.e. it was not designed keeping a cognitive secondary in mind. This constraint makes the classical interference channel \cite{car78} one sided, i.e, Z-interference channel with an additional power constraint. 
In this paper, we analyze how temporal opportunities due to sporadic primary traffic can be exploited, even if the secondary transmitter has incomplete information about the current state of primary traffic. 

Our contributions are three-fold. 
First, we approximate the uncertainty in primary activity (interference channel with a sporadic primary) by a simple block activity model where the primary changes its state at most once during a block of fixed duration $T+1$. In this model, the two sources of uncertainty in the primary traffic are captured by the initial state $s_0$ and the time of state change $\tau$. Their actual values are unknown to the secondary, but their distribution $\pi(s_{0})$ and $f_{T}(\tau)$ are known. We first show that the reliability constraint at the primary receiver, which requires that primary transmission should not be harmed, places an additional power constraint on the secondary transmitter (see for example, \cite{zm08}) and the interference caused by the sporadic primary at the secondary receiver converts the AWGN channel to a Gaussian mixture channel. 

Second, we present two sense-and-send schemes where the secondary senses the channel at the start of each block to look for temporal opportunities. The fixed primary design converts the effective channel into a MAC. The secondary splits its rate into two layers \cite{ru96}, \cite{py07}. We show that it is optimal to treat the primary message either as all public information or as all private information but not both because the fixed primary's message cannot be spilt into private and public parts. We also present a no-sensing scheme and a genie-aided upper bound. We prove that for the sense-and-send schemes the secondary power profile is paranoid, i.e. the power monotonically increases or decreases during a block depending on whether the primary was using the channel or not during the start of the block. The paranoid profile arises due to the effective noise of the mixture channels which decays or grows, depending on the starting state of the channel in a block. Additionally, we prove that when the sensing is perfect, the paranoid scheme is optimal. The proof is along the lines of a channel with delayed state information \cite{vis99}. 

Third, we show that if there is no information about the primary traffic the rate splitting scheme achieves capacity. The effective channel is converted to a MAC with a fixed transmitter from the secondary's point of view and hence rate splitting at the secondary transmitter and sequential decoding at the secondary receiver is optimal.  Finally, we show that the paranoid scheme approaches a genie-aided upper bound for large block lengths and when the primary traffic becomes more persistent, both these schemes get close to the no-information scheme. 

Much like any active area, many variations have been proposed and studied in the literature~\cite[and references therein]{gjm08}. Some of the earlier work \cite{mit00,jv06} in opportunistic spectrum allocation for cognitive flows was motivated by studies done by FCC \cite{fcc02} showing significant spectral underutilization. Many aspects of cognitive radio have been studied including coding with degraded messages \cite{dt06,wva07}, capacity of the secondary flow with causal or non-causal information \cite{js07}, stability of the queues at both the flows for maximal secondary rate with a guaranteed primary throughput \cite{sbs07}, spectral shaping \cite{zm08} etc. The case of an adaptive primary which acts as a feedback to the cognitive radio is considered in \cite{egr08}. When interference is considered in terms of SINR constraints (see for example, \cite{da08,gas07,gs07,gs06}), the effect of the primary communication at the secondary is not considered. Power control schemes have been derived for different power constraints \cite{zm08,zha08a} when the primary is persistent but operating below capacity. The idea of opportunistic interference cancellation was introduced in \cite{py07}. However all the above results do not address the issue of the uncertainty in the channel when the primary traffic is fixed but sporadic. We extend the idea of opportunistic interference cancellation to derive sense and send schemes for a fixed and sporadic primary operating below capacity.

The rest of the paper is organized as follows. In Section II, we present the system model and introduce the a block activity model for a sporadic primary. In Section III, we first present a genie-aided bound. Then we derive the paranoid secondary profile for a sense and send protocol when the sensing is noisy. Finally, we give two special cases when the sensing is noiseless and when no sensing is performed. We conclude with some numerical examples. 


\section{System Model \label{sec:model}}

\subsection{Signal Model \label{sec:signal}}

\begin{figure}[ht]
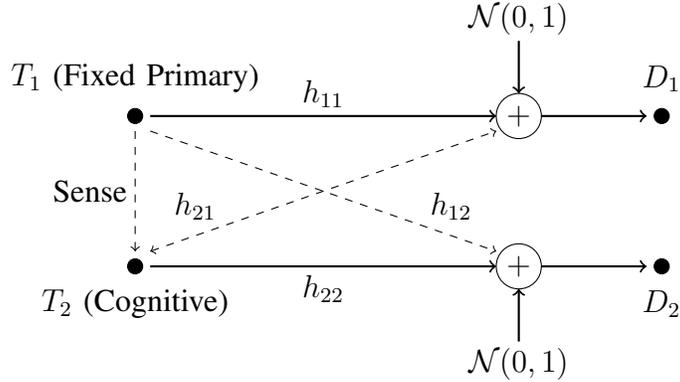

\centering
\beginpgfgraphicnamed{twoFlowBlock}
\tikz{
	\draw [fill=black] (1,4) circle (.1cm);
	\draw [fill=black] (8,4) circle (.1cm);
	\draw (6.1,4) circle (.3cm) node {$+$};
	\draw [fill=black] (1,2) circle (.1cm);
	\draw [fill=black] (8,2) circle (.1cm);
	\draw (6.1,2) circle (.3cm) node {$+$};
	\draw [->, thick] (1.2,4) -- (5.8,4);
	\draw [->, thick] (1.2,2) -- (5.8,2);
	\draw [->, dashed] (1.2,3.8) -- (5.8,2.2);
	\draw [<->, dashed] (1.2, 2.2) -- (5.8,3.8);
	\draw [->, dashed] (1,3.8) -- (1,2.2);
	\draw [->, thick] (6.1,5) -- (6.1,4.3);
	\draw [->, thick] (6.1,1) -- (6.1,1.7);
	\draw [->, thick] (6.4,4) -- (7.8,4);
	\draw [->, thick] (6.4,2) -- (7.8,2);
	\draw (1,4.5) node {$T_1$ (Fixed Primary)};
	\draw (8,4.5) node {$D_1$};
	\draw (8,1.5) node {$D_2$};
	\draw (1,1.5) node {$T_2$ (Cognitive)};
	\draw (0.4,3) node {Sense};
	\draw (6.1,5.3) node {$\mathcal{N}(0,1)$};
	\draw (6.1,0.7) node {$\mathcal{N}(0,1)$};
	\draw (5.2,2.8) node {$h_{12}$};
	\draw (1.8,2.8) node {$h_{21}$};
	\draw (3.5,4.3) node {$h_{11}$};
	\draw (3.5,1.7) node {$h_{22}$};
}
\endpgfgraphicnamed
\caption{ A two-flow cognitive interference network.
\label{fig:two-flow}}
\end{figure}

We consider an interference channel (Figure~\ref{fig:two-flow}), where the channel inputs and outputs are related as, $Y_1 = h_{11} X'_1  + h_{21} X'_2 + Z_1$ and  $Y_2 = h_{12} X'_1 + h_{22} X'_2 + Z_2,$
where $X'_1, X'_2$ are channel inputs, $Y_1,Y_2$ are the channel outputs and $Z_1, Z_2$ are independent zero mean, unit variance  Gaussian noise. 
Each transmitter is individually average power constrained, such that $E[X'^2_1]\leq P_1$ and $E[X'^2_2]\leq P_2$. 
We define the following for convenient interpretation:  $\text{SNR}_1=|h_{11}|^2P_1, \text{SNR}_2=|h_{22}|^2E[X'^2_2], \text{INR}_1=|h_{21}|^2E[X'^2_2], \text{INR}_2=|h_{12}|^2P_1$. We have fixed the primary power to $P_1$ (primary uses its full power budget) in our definition of $\text{SNR}_1$ and $\text{INR}_2$, leaving the secondary power as an optimization variable in our definition of $\text{SNR}_2$ and $\text{INR}_1$. 
Finally, we use the standard form \cite{cos85} of interference channel by assuming $|h_{11}|=|h_{22}|=1$. Thus the channel input-output are related as 
$Y_1 = \sqrt{\text{SNR}_1}X_1  + \sqrt{\text{INR}_1}X_2 + Z_1$ and 
$Y_2 =\sqrt{\text{INR}_2} X_1 + \sqrt{\text{SNR}_2}X_2 + Z_2,$
where the variance of $X_1$ and $X_2$ is bounded above by one. 

\subsection{Block Activity Traffic Model for the Primary Flow \label{sec:primary model}}

\begin{figure}[ht]
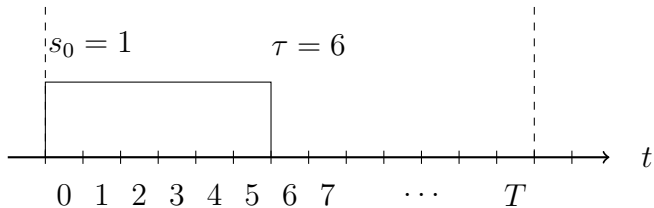

	\centering
	\tikz{
		\draw [->, thick] (-.5,0) -- (7.5,0);
		\draw[snake=ticks,segment length=.5 cm] (0,0) -- (7,0);
		\draw (8,0) node {$t$};
		\draw (.25,-.5) node {$0$};\draw (.75,-.5) node {$1$};
		\draw (1.25,-.5) node {$2$};\draw (1.75,-.5) node {$3$};
		\draw (2.25,-.5) node {$4$};\draw (2.75,-.5) node {$5$};
		\draw (3.25,-.5) node {$6$};\draw (3.75,-.5) node {$7$};
		\draw (5,-.5) node {$\ldots$};
		\draw (6.25,-.5) node {$T$};
		\draw (0,0) rectangle (3,1);
		\draw [dashed] (0,0) -- (0,2);
		\draw [dashed] (6.5,0) -- (6.5,2);
		\draw (.6,1.5) node {$s_0=1$};
		\draw (3.5,1.5) node {$\tau=6$};
	}
	\caption{Simplified block activity model for the primary packets with only one allowed switching of state. This model is characterized by the starting state $s_0$, the state-switching time $\tau$ and the blocklength $T+1$.
	\label{fig:blkModel}}
\end{figure}

The primary is assumed to have the following two properties: its channel is \emph{underutilized} i.e. it sends below the capacity of its own channel when it transmits, and its traffic is \emph{sporadic} i.e. it does not occupy the channel continuously.
We model the channel underutilization by assuming that the primary employs Gaussian random codes with a code rate of $ \mathcal{C} \left( \frac{\text{SNR}_{1}}{1+\text{INR}_\text{gap}} \right)$, where $\text{INR}_\text{gap} \geq 0$ and $\mathcal{C}(x)=\frac{1}{2}\log_{2}(1+x)$. The sporadic traffic is modeled by assuming that the primary transmits for $\beta$ fraction of time, which results in an average primary rate of $R_{1}=\beta\mathcal{C} \left( \frac{\text{SNR}_{1}}{1+\text{INR}_\text{gap}} \right)$. Thus a secondary user can use the same channel as long as it ensures that the primary INR does not go above  $\text{INR}_\text{gap}$. We refer to this as the \emph{INR constraint}.
We capture the two uncertainties from the secondary's point of view, the start time of the primary transmission bursts and the duration of the bursts, using the following block activity model. The primary transmissions are assumed to occur in blocks of $T+1$ time-slots, where $T$ is a finite constant. All transmissions by the primary and the secondary are considered to be slot synchronous. The primary user activity is labeled as the state of the primary channel and is denoted by $s_t$, where $0\leq t\leq T$ is the time-slot index in a block. The state process is independent of the secondary channel's input and output. The primary channel is either in the busy state, $s_t=1$, or in the idle state, $s_t=0$ during the time-slot $t$.
To keep the analysis tractable, we assume that the primary user changes its state only once at time-slot $\tau$ in a block (see Figure~\ref{fig:blkModel}). Thus, the tuple $(s_0,\tau)$ captures the two uncertainties related to the sporadic transmissions.

The starting state of the primary user for each block is assumed to be independently drawn from $\pi(s_0)= [\pi(0), \pi(1)]$. Conditioned on the starting state, the switching time of the primary $\tau$ has a probability mass function, $f_T(\tau), 1\leq \tau \leq T+1$ and the corresponding distribution function $F_T(\tau)=\sum\limits_{i=1}^{\tau}f_T(i)$. If $\tau=T+1$, then the primary user does not change its state during the block.
The secondary transmitter is assumed to know $\text{INR}_{\text{gap}}$, $\pi(s_{0})$ and $f_{T}(t)$. The secondary receiver is assumed to have perfect knowledge to do coherent decoding.

 Since the primary uses Gaussian codes, from the point of view of the secondary the channel always behaves like an AWGN channel. The effective noise as seen by the secondary receiver has the mixture distribution for $1\leq t\leq T$,
\begin{equation}
Z^{(t)} = \begin{cases}
\mathcal{N}(0,1) & \mbox{ if $s_t=0$ (probability $=\bar{\beta}_s(t)$)} \\ 
\mathcal{N}(0,1+\text{INR}_2) & \mbox{ if $s_t=1$ (probability $=\beta_s(t)$)}\\
\end{cases},
\label{eqn:effectiveNoise}
\end{equation} 
where, $\bar{\beta}_s(t)=1-\beta_s(t)$. The term $\beta_{s}(t)$ is the probability of the state $s_{t}=1$ conditioned on the starting state $s$, and is used to derive the power profile in Section~III. Note that the $\beta$ is the actual primary channel usage fraction and $\beta_{s}(t)$ is what is seen by the secondary.
A persistent primary can be represented by putting $\beta=1$. If $\text{INR}_\text{gap} =0$, the system is fully loaded and no secondary is allowed on the same channel. The interesting case is when $\text{INR}_\text{gap} >0$ and $\beta < 1$. 

\subsection{Interleaved Block Code for the Secondary Flow\label{sec:codeD}}  
 
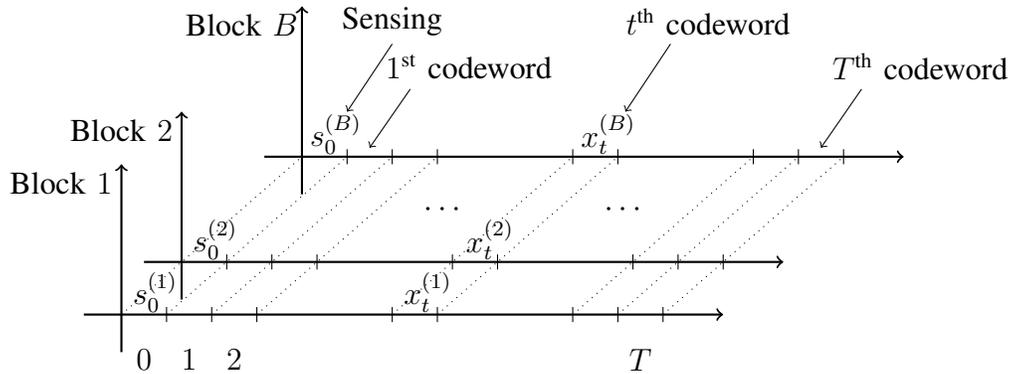
\begin{figure}[ht]
\centering
\beginpgfgraphicnamed{interleavedCoding}
\begin{tikzpicture}
	\def \ticstep {0.6};
	\def \Xaxisstep {.8};
	\def \Yaxisstep {.7};
	\def \xlen {8};
	\def \ylen {2};
	%
	\def \xz {0};
	\def \yz {0};
	\draw [->, thick] (\xz-0.5,\yz) -- (\xlen+\xz,\yz); 
	\draw [->, thick] (\xz,\yz-0.5) -- (\xz,\ylen+\yz); 
	\draw (\xz+\ticstep, \yz-.1) -- (\xz+\ticstep,\yz+0.1);
	\draw (\xz+2*\ticstep,\yz-0.1) -- (\xz+2*\ticstep,\yz+0.1);
	\draw (\xz+3*\ticstep,\yz-0.1) -- (\xz+3*\ticstep,\yz+0.1);
	\draw (\xz+6*\ticstep,\yz-0.1) -- (\xz+6*\ticstep,\yz+0.1);
	\draw (\xz+7*\ticstep,\yz-0.1) -- (\xz+7*\ticstep,\yz+0.1);
	\draw (\xz+10*\ticstep,\yz-0.1) -- (\xz+10*\ticstep,\yz+0.1);
	\draw (\xz+11*\ticstep,\yz-0.1) -- (\xz+11*\ticstep,\yz+0.1);
	\draw (\xz+12*\ticstep,\yz-0.1) -- (\xz+12*\ticstep,\yz+0.1);
         %
	\def \xz {\Xaxisstep};
	\def \yz {\Yaxisstep};
	\draw [->, thick] (\xz-0.5,\yz) -- (\xlen+\xz,\yz); 
	\draw [->, thick] (\xz,\yz-0.5) -- (\xz,\ylen+\yz); 
	\draw (\xz+\ticstep, \yz-.1) -- (\xz+\ticstep,\yz+0.1);
	\draw (\xz+2*\ticstep,\yz-0.1) -- (\xz+2*\ticstep,\yz+0.1);
	\draw (\xz+3*\ticstep,\yz-0.1) -- (\xz+3*\ticstep,\yz+0.1);
	\draw (\xz+6*\ticstep,\yz-0.1) -- (\xz+6*\ticstep,\yz+0.1);
	\draw (\xz+7*\ticstep,\yz-0.1) -- (\xz+7*\ticstep,\yz+0.1);
	\draw (\xz+10*\ticstep,\yz-0.1) -- (\xz+10*\ticstep,\yz+0.1);
	\draw (\xz+11*\ticstep,\yz-0.1) -- (\xz+11*\ticstep,\yz+0.1);
	\draw (\xz+12*\ticstep,\yz-0.1) -- (\xz+12*\ticstep,\yz+0.1);
	%
	\def \xz {3*\Xaxisstep};
	\def \yz {3*\Yaxisstep};
	\draw [->, thick] (\xz-0.5,\yz) -- (\xlen+\xz,\yz); 
	\draw [->, thick] (\xz,\yz-0.5) -- (\xz,\ylen+\yz); 
	\draw (\xz+\ticstep, \yz-.1) -- (\xz+\ticstep,\yz+0.1);
	\draw (\xz+2*\ticstep,\yz-0.1) -- (\xz+2*\ticstep,\yz+0.1);
	\draw (\xz+3*\ticstep,\yz-0.1) -- (\xz+3*\ticstep,\yz+0.1);
	\draw (\xz+6*\ticstep,\yz-0.1) -- (\xz+6*\ticstep,\yz+0.1);
	\draw (\xz+7*\ticstep,\yz-0.1) -- (\xz+7*\ticstep,\yz+0.1);
	\draw (\xz+10*\ticstep,\yz-0.1) -- (\xz+10*\ticstep,\yz+0.1);
	\draw (\xz+11*\ticstep,\yz-0.1) -- (\xz+11*\ticstep,\yz+0.1);
	\draw (\xz+12*\ticstep,\yz-0.1) -- (\xz+12*\ticstep,\yz+0.1);
	%
	\draw [dotted] (0,0) -- (\xz,\yz); 
	\draw [dotted] (\ticstep,0) -- (\xz+\ticstep,\yz); 
	\draw [dotted] (2*\ticstep,0) -- (\xz+2*\ticstep,\yz); 
	\draw [dotted] (3*\ticstep,0) -- (\xz+3*\ticstep,\yz); 
	\draw [dotted] (6*\ticstep,0) -- (\xz+6*\ticstep,\yz); 
	\draw [dotted] (7*\ticstep,0) -- (\xz+7*\ticstep,\yz); 
	\draw [dotted] (10*\ticstep,0) -- (\xz+10*\ticstep,\yz); 
	\draw [dotted] (11*\ticstep,0) -- (\xz+11*\ticstep,\yz); 
	\draw [dotted] (12*\ticstep,0) -- (\xz+12*\ticstep,\yz); 
	\draw (-0.8,\ylen-0.25) node {Block $1$};
	\draw (\Xaxisstep-0.8,\Yaxisstep+\ylen-0.25) node {Block $2$};
	\draw (\xz-0.8,\yz+\ylen-0.25) node {Block $B$};
	\draw (\ticstep/2,-\ticstep) node {$0$};
	\draw (\ticstep/2+\ticstep,-\ticstep) node {$1$};
	\draw (\ticstep/2+2*\ticstep,-\ticstep) node {$2$};
	\draw (\ticstep/2+11*\ticstep,-\ticstep) node {$T$};
	\draw (2*\Xaxisstep+4.5*\ticstep,2*\Yaxisstep) node {$\cdots$};
	\draw (2*\Xaxisstep+8.5*\ticstep,2*\Yaxisstep) node {$\cdots$};
	\draw (3*\ticstep/4,\ticstep/2) node {$s_0^{(1)}$};
	\draw (\Xaxisstep+3*\ticstep/4,\Yaxisstep+\ticstep/2) node {$s_0^{(2)}$};
	\draw (\xz+3*\ticstep/4,\yz+\ticstep/2) node {$s_0^{(B)}$};
	\draw (2*\Xaxisstep+4*\ticstep+\ticstep/8,\ticstep/2) node {$x_t^{(1)}$};
	\draw (2*\Xaxisstep+5*\ticstep+\ticstep/2,\Yaxisstep+\ticstep/2) node {$x_t^{(2)}$};
	\draw (2*\Xaxisstep+8*\ticstep+\ticstep/8,\yz+\ticstep/2) node {$x_t^{(B)}$};
	\draw (\xz+2*\ticstep, \yz+3*\ticstep) node {Sensing};
	\draw [->] (\xz+2*\ticstep,\yz+2.5*\ticstep) -- (\xz+\ticstep,\yz+\ticstep);
	\draw (\xz+3.7*\ticstep,\yz+2*\ticstep) node {$1^{\text{st}}$ codeword};
	\draw [->] (\xz+2.4*\ticstep,\yz+1.5*\ticstep) -- (\xz+1.5*\ticstep,\yz+.2*\ticstep);
	\draw (\xz+13.7*\ticstep,\yz+2*\ticstep) node {$T^{\text{th}}$ codeword};
	\draw [->] (\xz+12.4*\ticstep,\yz+1.5*\ticstep) -- (\xz+11.5*\ticstep,\yz+.2*\ticstep);
	\draw (\xz+9*\ticstep,\yz+3*\ticstep) node {$t^{\text{th}}$ codeword};
	\draw [->] (2*\Xaxisstep+9.5*\ticstep,\yz+2.5*\ticstep) -- (2*\Xaxisstep+8*\ticstep+\ticstep/2,\yz+\ticstep);
\end{tikzpicture}
\endpgfgraphicnamed
\caption{Codewords are multiplexed across $B$ blocks, i.e. each time slot uses a different codeword which spans across $B$ blocks. The figure shows the sensing operation at the beginning of each block and the $t^{\text{th}}$ codeword, $(x_{t}^{(1)},\ldots,x_{t}^{(B)})$. Each of the $T$ codewords are picked from the corresponding $T$ codebooks. The block size $T$ is a positive constant, however, the code length $B$ tends to infinity. 
\label{fig:multiplexedCW}}
\end{figure}

We now define the set of secondary codes  over $B$ consecutive blocks, when the secondary has an estimate of the starting state $\hat{s}_0$ for each block. Bold face letters represent the vector corresponding to the variable for all blocks, e.g. $\mathbf{s}_{0}=(s_{0}^{(1)},s_{0}^{(2)}\ldots,s_{0}^{(B)})$ represents the starting state of all $B$ blocks. 
 Since the $t^{\text{th}}$ time-slot of each block has identically distributed noise as given by Equation~(\ref{eqn:effectiveNoise}), 
 the secondary designs $T$ different codebooks each matched to a given time slot across $B$ blocks. For example, during $b^{\text{th}}$ block the secondary transmits $(x_{1}^{(b)},x_{2}^{(b)},\ldots,x_{T}^{(b)})$ where $x_{j}^{(i)}$ denotes the $i^{\text{th}}$ component of a codeword from the $j^{\text{th}}$ codebook. Similarly the $t^{\text{th}}$ codeword is given by $\mathbf{x_{t}}=(x_{t}^{(1)},x_{t}^{(2)}\ldots,x_{t}^{(B)})$ as shown in Figure~\ref{fig:multiplexedCW}.

A cognitive interleaved block code of length $N=BT$ consists of $T$ separate $(B,2^{B(R-\epsilon)},\epsilon)$-codes, corresponding to each time slot of a block. Each of these codebooks  further consists of two component codebooks corresponding to the sensed starting state of the block $\hat{s}_0$ which may or may not match the actual starting state $s_{0}$. The codebooks are defined by a set of $2T$ encoding functions that map the set of equiprobable messages $\{ 1,\ldots,2^{BR}\}$ to channel inputs. The transmitter encoding function is defined as $f_{t\hat{s}_{0}} : \mathcal{W} \times \mathcal{S} \rightarrow \mathcal{X}$ for $t=1,\ldots,T$ and $\hat{s}_{0} =1,2$, i.e. $\mathbf{x}_{t\hat{s}_{0}}=f_{t\hat{s}_{0}}(w,\hat{s}_0)$.
The decoding function $g:\mathcal{Y}^{B} \times \mathcal{S}^{B} \rightarrow \mathcal{W}$ maps the received vector to the message set, leading to probability of error  
$P_e = \frac{1}{2^{BR}} \sum\limits_{i=1}^{2^{BR}} \sum\limits_{\mathbf{s}_{0}} p(\mathbf{s}_0) \sum\limits_{\mathbf{y}: g(\mathbf{y},\mathbf{s}_0)\neq i} P_e(\mathbf{y}|\mathbf{f}(i),\mathbf{s}_0).$ 
The rate $R$ is $\epsilon\text{-achievable}$ if there exists a cognitive block code for sufficiently large $N=BT$ and $P_{e}\leq \epsilon$. A rate $R$ is said to be achievable if there exists an $\epsilon\text{-achievable}$ code for every $\epsilon > 0$ and the capacity is defined to be the supremum of all the achievable rates $R$.

Finally consider the Han-Kobayashi scheme for interference channels where each user's data is split into private and public parts. The public information is decodable by both decoders and the private information is decodable by the intended decoder. A special case of this structure is usable in our model. Firstly, since the primary encoder is fixed, its data can be considered either public (if the secondary can decode it, i.e. the cross channel can support the rate) or private, but it cannot be split into both. Secondly, since the primary decoder is fixed, it considers all the secondary information as noise, i.e. private. Since the primary data is fixed, the secondary can do rate splitting at the encoder and sequential decoding at the decoder without losing optimality. We will call the first layer the single user codeword with power $\rho_{\hat{s}Nt}$ since it is decoded using a single user decoder (treating everything else as \emph{Noise}), and the second layer the multiuser codeword with power $\rho_{\hat{s}St}$ since it is decoded using a multiuser receiver (with \emph{Successive} interference cancellation). The subscript $s,\hat{s}$ will be replaced by its value (0 or 1) depending on the context.

\section{Main Results \label{sec:main}}

In the block activity model, the two sources of uncertainty in the primary traffic are captured by the initial state $s_0$ and the time of state change $\tau$. Their actual values are unknown to the secondary, but their distribution $f_{T}(\tau)$ and $\pi(s_{0})$ are assumed to be known.  Our aim is to understand how lack of knowledge of these two parameters impacts the secondary rate.
We first derive the case where a Genie provides the information about $(s_0,\tau)$ which serves as an upper bound and a design motivation for the general case of Section~\ref{sec:nsysense}, where the secondary has an estimate $\hat{s}_0$ of the initial state $s_0$ and has no knowledge of $\tau$. We then consider two important special cases of the general scheme. First, we consider the case of perfect sensing, $\hat{s}_0 = s_0$, and derive stronger results about the optimal secondary transmission design. Second, we consider the case where the secondary has no information about any of the unknown parameters, which is equivalent to $\hat{s}_0$ not providing any useful information about $s_0$. 

\subsection{Genie-aided Case: Secondary has perfect estimate of $(s_{0},\tau)$\label{sec:genie}}

In this section, we assume Genie-aided knowledge of  $(s_{0},\tau)$ at the secondary. As described in Section~\ref{sec:primary model}, the primary flow has two states, on and off. Hence the secondary transmitter uses two interleaved codes matched to the two channel states. Such a solution is called water-filling or water-pouring \cite{gal68}, as the variance of the two codebooks is inversely proportional to the noise variance of the corresponding channel state.  
We use the following result from \cite{lap96} to show that Gaussian codebooks are optimal for the secondary.
\begin{lemma} [Optimality of Gaussian codewords, see {\cite [Theorem~$1$]{lap96}}] For a single-user scalar additive noise channel with nearest neighbor decoding irrespective of the noise distribution, the average probability of error over the ensemble of Gaussian codebooks of power $P$, approaches zero as the blocklength $n$ tends to infinity for code rates below $\log (1+P/N)$ (and approaches one for rates above $\log (1+P/N)$).
\label{lem:Gopt}
\end{lemma}
 Since the primary uses a Gaussian codebook, the equivalent noise at the secondary is also Gaussian.  Further since the decoder of the primary is fixed, any distribution of noise at its receiver does not change the primary capacity as long as the variance (interference plus noise) is below $\text{INR}_{\text{gap}}$. Hence Gaussian codebooks are optimal for the secondary. The secondary uses a Gaussian codebook with power $\rho_0$ when the primary is off, and a superposition Gaussian codebook with power $(\rho_{1N},\rho_{1S})$. Define $\text{INR}_\text{C}=\frac{\text{INR}_\text{gap}}{|h_{21}|^2}$ and $\text{SIC}_\text{C}=|h_{12}|^2(\text{INR}_\text{gap}+1)-1$ to characterize the power allocation for the Gaussian codebook that achieves capacity.
\begin{theorem} [Genie-aided Upper Bound]
When $\text{SIC}_{\text{C}}\geq \min\left( \text{SNR}_{2},\text{INR}_{\text{C}}\right)$, then the optimal power allocation $(\rho_0,\rho_{1N},\rho_{1S}) $ for the genie-aided case is given by
\begin{equation}
\left\{ \begin{array}{ll} 
(\text{SNR}_{2},0,\text{SNR}_{2}) & \mbox{if $\text{SNR}_{2} \leq \text{INR}_{\text{C}}$} \\ 
(\frac{\text{SNR}_{2}-\beta\text{INR}_{\text{C}}}{\bar{\beta}}, 0, \text{INR}_\text{C}) & \mbox{if $\text{INR}_{\text{C}} \leq \text{SNR}_{2}$} 
\end{array},
\right.
\label{eqn:genie1}
\end{equation} 
On the other hand, when $\text{SIC}_\text{C} < \min \left( \text{SNR}_2, \text{INR}_\text{C} \right)$, then the optimal power allocation $(\rho_0,\rho_{1N},\rho_{1S}) $ is given by,
\begin{equation}
\left\{ \begin{array}{ll} 
(\frac{\gamma}{\bar{\beta}}, 0, \text{SIC}_\text{C}) & \mbox{if $\text{SIC}_\text{C}\leq \text{SNR}_2\leq \text{SIC}'_\text{C}$} \\ 
(\gamma+\beta\alpha, \gamma-\bar{\beta}\alpha, \text{SIC}_\text{C}) & \mbox{if $\text{SIC}'_\text{C}\leq \text{SNR}_2\leq \text{SIC}'_\text{C}  +\text{INR}_\text{C}-\text{SIC}_\text{C}$} \\ 
(\frac{\gamma-\beta\delta}{\bar{\beta}}, \delta, \text{SIC}_\text{C}) & \mbox{if $\text{SNR}_2\geq \text{SIC}'_\text{C} +\text{INR}_\text{C}-\text{SIC}_\text{C}$} 
\end{array},
\right.
\label{eqn:genie2}
\end{equation}
where 
$\text{SIC}'_\text{C}=\text{SIC}_\text{C} +\beta \text{INR}_2$, 
$\alpha=\text{SIC}_\text{C}+\text{INR}_2$, 
$\gamma=\text{SNR}_2-\beta \text{SIC}_\text{C} \text{ and } \delta=\text{INR}_\text{C}-\text{SIC}_\text{C}$. Here, $\beta$ and $\bar{\beta}=1-\beta$ are the on and off fractions of the primary, respectively.
\label{thm:genie}
\end{theorem}

\begin{proof} The proof is based on two steps. First we use Lemma~\ref{lem:Gopt} to show that Gaussian codebooks are optimal for the secondary transmissions. Then the optimal power allocation is obtained by solving the following optimization problem,
\begin{eqnarray}
&\max\limits_{\rho_0,\rho_{1N},\rho_{1S}\geq 0} &\bar{\beta} \mathcal{C}(\rho_0)+\beta\left[\mathcal{C}\left( \frac{\rho_{1N}}{1+\text{INR}_2+\rho_{1S}} \right)+\mathcal{C}(\rho_{1S})\right]\\
&\text{subject to }&\bar{\beta} \rho_0+\beta(\rho_{1N}+\rho_{1S}) \leq \text{SNR}_2\notag\\
&&\rho_{1N}+\rho_{1S} \leq \frac{\text{INR}_\text{gap}}{|h_{21}|^2}\notag\\
&&\rho_{1S} \leq |h_{12}|^2(1+\text{INR}_\text{gap})-1.\notag
\label{eqn:genieCon}
\end{eqnarray}
The rate achieved is a corner point of the 3 user virtual MAC formed by the primary and the two code layers of the secondary. Hence the sequential decoding at the secondary receiver achieves capacity. Complete details are in Appendix~\ref{proof:genie}
\end{proof}

\begin{figure}[ht]
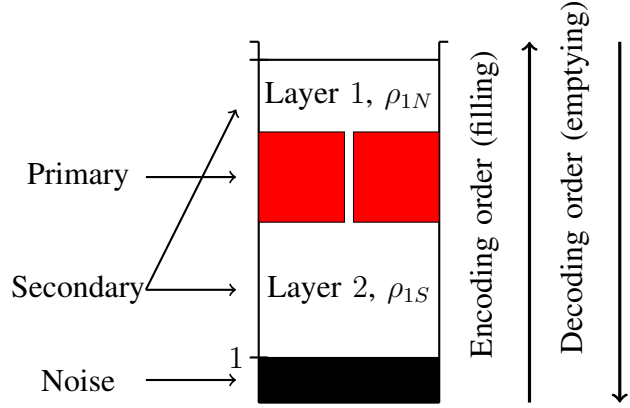

	\centering
	\tikz[scale=1.2]{
		\draw [thick] (2,0) -- (2,4);
		\draw [thick] (2,0) -- (4,0);
		\draw [thick] (4,0) -- (4,4);
		\draw [thick] (1.9,4) -- (2,4);
		\draw [thick] (4,4) -- (4.1,4);
		\draw [thick] (1.9,0.5) -- (2.1,0.5);
		\draw (1.75,0.5) node {$1$};
		\draw [fill=black] (2,0) rectangle (4,0.5);
		\draw [fill=red] (2,2) rectangle (2.95, 3); 
		\draw [fill=red] (3.05,2) rectangle (4,3);
		\draw [thick] (2,3.8) -- (4,3.8);
		\draw [thick] (1.9,3.8) -- (2.1,3.8);
		\draw (3,3.4) node {Layer $1$, $\rho_{1N}$};
		\draw (3,1.25) node {Layer $2$, $\rho_{1S}$};
		\draw [<-, very thick] (5,4) -- (5,0);
		\draw (4.5,2.2) node[rotate=90] {Encoding order (filling)};
		\draw [<-, very thick] (6,0) -- (6,4); 
		\draw (5.5,2.5) node[rotate=90] {Decoding order (emptying)};
		\draw [->, thick] (0.75,0.25) -- (1.75,0.25);
		\draw (0,0.25) node {Noise};
		\draw [->, thick] (0.75,1.25) -- (1.75,1.25);
		\draw [->, thick] (0.75,1.25) -- (1.75,3.25);
		\draw (0,1.25) node {Secondary};
		\draw [->, thick] (0.75,2.5) -- (1.75,2.5);
		\draw (0,2.5) node {Primary};
	}
	\caption{ Layered water-filling: power allocation can be thought of as filling a layered bucket with water. The presence of the primary causes one of the layers to be fixed. The size of the fixed layer (in red) is a function of $\text{INR}_2$, the size of Layer~$2$ depends on the interference gap $\text{INR}_{\text{gap}}$ and the rest of the power is put into Layer~$1$. When the primary is persistent, there is only one bucket to fill, as shown here. For a sporadic primary, there is another bucket without any layers for $\rho_0$. For the block activity model, each of the time slots across the block look alike with $(\rho_N, \rho_S)$ replaced by $(\rho_{0Nt}, \rho_{0St})$ or $(\rho_{1Nt}, \rho_{1St})$ depending on the starting state. 
	\label{fig:layered-water-filling}}
\end{figure}

As shown in Figure~\ref{fig:layered-water-filling}, power allocation can be viewed as water-filling with a layer in the middle representing the active primary. 
For the time-slots when primary is off, the secondary uses a single-user codebook matched to its own channel capacity. When the primary is on, the secondary transmitter splits into a two-layer Gaussian superposition codebook as shown in Figure~\ref{fig:layered-water-filling} followed by a sequential decoding at the secondary receiver. 
The sequence of decoding at the secondary receiver is as follows. First, Layer~$1$ is decoded by treating everything else as noise, then the primary codeword is decoded by treating Layer~2 as noise and finally Layer~$2$ is decoded interference-free. 

%

\subsection{ The General Case: Secondary with noisy estimate of primary state $\hat{s}_{0}$ \label{sec:nsysense}}
When the secondary does not know the channel states $s_{t}$ exactly for each time slot, it senses the channel in the beginning of each block $\hat{s}_{0}$ (noisy estimate of $s_{0}$). The effective channel is a Gaussian mixture channel as given by Equation~(\ref{eqn:effectiveNoise}). A given time-slot (for the same sensed state) has the same channel statistics across different blocks as shown in Figure~\ref{fig:multiplexedCW}.
We take a cue from the form of Genie-aided code in Theorem~\ref{thm:genie} and use a general superposition code. 
If $\hat{s}_0=0$, the secondary transmitter sends at a power level $\bm{\rho}_0=\bm{\rho}_{0N} + \bm{\rho}_{0S}$, (where $\bm{\rho}_{0S} = (\rho_{0S1},\ldots,\rho_{0ST})$ etc.) and if $\hat{s}_0=1$, it sends at a power level $\bm{\rho}_1=\bm{\rho}_{1N} + \bm{\rho}_{1S}$. 
The above time-dependent power allocation exploits secondary's knowledge of the conditional probability distribution of effective noise in each time slot caused by the primary. 
We shall omit the subscript $0$ in $s_{0}, \hat{s}_{0}$ in the summation indices to avoid clutter.
The error in the state estimate is characterized by the probability of missed detection $\mathcal{P}_M=P(\hat{s}_0=0|s_0=1)$ and the probability of false alarm $\mathcal{P}_F=P(\hat{s}_0=1|s_0=0)$.
The exact capacity with noisy state estimate $\hat{s}_0$ is unknown. To derive an achievable rate using Gaussian codes for secondary transmissions, we extend the optimization problem in~(\ref{eqn:genieCon}) by generalizing the rate and the three constraints for each sensed state.  

The average power constraint can be computed by summing the power levels weighted with the appropriate probability of occurrences,
\begin{align}
& \frac{\pi(0) (1-\mathcal{P}_F)}{T+1}\sum\limits_{t=1}^{T} (\rho_{0Nt}+\rho_{0St}) + \frac{\pi(0)\mathcal{P}_F}{T+1}\sum\limits_{t=1}^{T} (\rho_{1Nt}+\rho_{1St}) \notag\\
&\qquad  + \frac{\pi(1) \mathcal{P}_M}{T+1}\sum\limits_{t=1}^{T} (\rho_{0Nt}+\rho_{0St}) + \frac{\pi(1) (1-\mathcal{P}_M)}{T+1}\sum\limits_{t=1}^{T} (\rho_{1Nt}+\rho_{1St})
 \leq \text{SNR}_2 \notag\\ 
&\text{i.e. } \sum\limits_{s\in\{0,1\}}\pi(s)\sum\limits_{\hat{s}\in\{0,1\}}P(\hat{s}|s)\sum\limits_{t=1}^{T}(\rho_{\hat{s}Nt}+\rho_{\hat{s}St}) 
 \leq (T+1)\text{SNR}_2. \label{eqn:avgP}
\end{align}

Similarly the rate has to be calculated for the four possible combinations of $(s_0,\hat{s}_0)$. When the primary user is switched `off' or `on,' the effective secondary channel is AWGN with noise $\mathcal{N}(0,1)$ or $\mathcal{N}(0,1+\text{INR}_{2})$ respectively. When there is no primary on the channel, the secondary can support a rate of $\mathcal{C}\left(\frac{\rho_{0Ni}}{1+\rho_{0Si}}\right)+\mathcal{C}(\rho_{0Si})=\mathcal{C}(\rho_{0Si}+\rho_{0Ni})$ and in the presence of the primary, the secondary can support a rate of $\mathcal{C}\left(\frac{\rho_{0Ni}}{1+\text{INR}_2+\rho_{0Si}}\right)+\mathcal{C}(\rho_{0Si})$. 

\paragraph{$s_0=0, \hat{s}_0=0$}
When $s_0=0, \hat{s}_0=0$, the secondary correctly detects that the primary is silent and sends $\bm{\rho}_0=(\bm{\rho}_{0N}, \bm{\rho}_{0S})$. 
The achievable rate averaged over all possible primary switches is given by,
\begin{align*}
R_{00}
= {} & \sum\limits_{t=1}^{T} \left[ 
       f_T(t)\left(\sum\limits_{i=1}^{t-1}\mathcal{C}(\rho_{0Si}+\rho_{0Ni}) 
       + \sum\limits_{i=t}^{T}\left(\mathcal{C}\left(\frac{\rho_{0Ni}}{1+\text{INR}_2+\rho_{0Si}}\right)+\mathcal{C}(\rho_{0Si})\right)\right) \right]\\
\stackrel{(a)}{=} {} & \sum\limits_{t=1}^{T}\left[(1-F_T(t))\mathcal{C}(\rho_{0St}+\rho_{0Nt}) 
       +  F_T(t) \left(\mathcal{C}\left(\frac{\rho_{0Nt}}{1+\text{INR}_2+\rho_{0St}}\right)+\mathcal{C}(\rho_{0St})\right) \right]\\
= {} & \sum\limits_{t=1}^{T} \left( \bar{\beta}_0(t)\mathcal{C}_{\hat{s}0} + \beta_0(t)\mathcal{C}_{\hat{s}1} \right),
\end{align*}
where $\mathcal{C}_{\hat{s}0} = \mathcal{C}(\rho_{\hat{s}St}+\rho_{\hat{s}Nt})$, $\mathcal{C}_{\hat{s}1} = \left( \mathcal{C}\left(\frac{\rho_{\hat{s}Nt}}{1+\text{INR}_2+\rho_{\hat{s}St}}\right)+\mathcal{C}(\rho_{\hat{s}St}) \right)$, $\bar{\beta}_0(t)=1-F_T(t)$ and $\beta_0(t)=F_T(t)$. 
Lemma~\ref{lem:chOrder} (see Appendix~\ref{summChange}) was used to change the summation order in step (a).

\paragraph{$s_0=1, \hat{s}_0=1$}

In this case, as the secondary correctly detects the primary user and sends $\bm{\rho}_0=(\bm{\rho}_{0N}, \bm{\rho}_{0S})$ and the effective noise at the secondary receiver  has a variance of $1+\text{INR}_2$ in the beginning of the block which changes to $1$ when the primary user changes its state at time-slot $\tau=t$. Probability of this event $s=1$, is  $\pi(1)$. The rate that can be achieved in this case is given by,
$R_{11} =  \sum\limits_{t=1}^{T} ( \beta_1(t)\mathcal{C}_{\hat{s}1}+\bar{\beta}_1(t)\mathcal{C}_{\hat{s}0} )$,
where $\bar{\beta}_1(t)=F_T(t)$ and  $\beta_1(t)=1-F_T(t)$.

\paragraph{$s_0=1, \hat{s}_0=0$}

Even though the primary user is transmitting, the secondary detects that there is no primary packet on the channel, it sends $(\bm{\rho}_{0N},\bm{\rho}_{0S})$. In this case the primary user is actually on during the start of the block, the noise that the secondary receiver sees has a variance of $1+|h_{12}|^2\text{SNR}_1$ and it changes to $1$ when the primary changes its state. Probability of the event $(s_0=1, \hat{s}_0=0)$, is  $\pi(1)\mathcal{P}_M$. The rate achieved for a given $t$ this case is given by, 
$R_{10} = \sum\limits_{t=1}^{T} ( \beta_1(t)\mathcal{C}_{\hat{s}1} + \bar{\beta}_1(t)\mathcal{C}_{\hat{s}0} ).$

\paragraph{$s_0=0, \hat{s}_0=1$}

In this case, as the secondary detects the primary user and it sends $(\rho_{11},\ldots,\rho_{1T})$, but the primary user is actually off during the start of the block, so the noise that the secondary receiver sees in the beginning is $N_2$ and it changes to $P_1+N_2$ when the primary changes its state. Probability of the event $(s_0=0, \hat{s}_0=1)$ is  $\pi(0)\mathcal{P}_F$. The rate achieved for a given $t$ is given by, 
$R_{01} = \sum\limits_{t=1}^{T} ( \bar{\beta}_0(t)\mathcal{C}_{\hat{s}0} +\beta_0(t)\mathcal{C}_{\hat{s}1} ).$

Adding up all the terms derived above after weighting them with the appropriate probability of occurrences, leads to the average rate of
$R_{2}(\bm{\rho})
=\frac{1}{2(T+1)} [\pi(0)(1-\mathcal{P}_F) R_{00} +\pi(1)(1-\mathcal{P}_M) R_{11}+ \pi(1)\mathcal{P}_M R_{10}+\pi(0)\mathcal{P}_FR_{01} ].$

Two additional constraints are required to complete the problem formulation. 
First the \emph{INR constraint} which is imposed due to the constraint on the maximum noise variance that can be experienced at the primary receiver, $\rho_{\hat{s}Nt}+\rho_{\hat{s}St}\leq \text{INR}_\text{C}$, for $t \in [1, T]$ and  $s \in \{0,1\}$.
Second, the \emph{SIC constraint} which is imposed to ensure that the primary information can be decoded in the presence of the multiuser codeword $\rho_{\hat{s}St}$, after the single user codeword $\rho_{\hat{s}Nt}$ has been decoded out. This gives rise to the condition, $\mathcal{C}\left(\frac{\text{SNR}_1}{1+\text{INR}_\text{gap}}\right) \leq \mathcal{C}\left(\frac{\text{INR}_2}{1+\rho_{sSt}}\right)$ which is same as $\rho_{sSt} \leq \text{SIC}_\text{C}$,  for $t \in [1, T]$ and  $s \in \{0,1\}$. Note that these two constraints hold for each state and time slot in contrast to the average power constraint which are an average constraint. 
We solve the following optimization problem, to find the optimal power profile,
\begin{eqnarray}
&\max\limits_{\bm{\rho} \in \Phi}
&R_{2}(\bm{\rho})= 
 \frac{1}{2(T+1)} \sum\limits_{s} \pi(s) \sum\limits_{\hat{s}} P(\hat{s}|s)\sum\limits_{t=1}^{T} ( \bar{\beta}_s(t)\mathcal{C}_{\hat{s}0} +\beta_s(t)\mathcal{C}_{\hat{s}1} ),
\label{eqn:nsenseOptProb}
\end{eqnarray}
where $\Phi$ is the constraint set defined by the INR, SIC and average power constraints derived above. Additionally, we have to consider positivity constraints for all the power variables, i.e. $\rho_{\hat{s}Nt}, \rho_{\hat{s}St} \geq 0 \text{ for } \hat{s}\in \{0,1\}, t\in [1,T]$.
This is with the understanding that the constraints do not become infeasible, i.e. $T, \text{INR}_\text{gap}, \text{SNR}_2 \geq 0$ and if $|h_{12}|^2(\text{INR}_\text{gap}+1)\leq 1$, $\rho_{sSt}=0$. We will assume complementarity and positivity but not discuss it further due to lack of space.

\begin{theorem}[Monotonicity of the power profile]
For the optimization problem given in (\ref{eqn:nsenseOptProb}), 
$\rho^*_{0N1} \geq \rho^*_{0N2} \geq \ldots \geq \rho^*_{0NT} \text{ and } \rho^*_{1N1} \leq \rho^*_{1N2} \leq \ldots \leq \rho^*_{1NT}.$ Additionally, 
\begin{enumerate}
\item[(i)] If $\rho^*_{0Nt}=0$ then $\rho^*_{0Nt+1},\ldots,\rho^*_{0NT}=0$
\item[(ii)] If $\rho^*_{1Nt}=0$ then $\rho^*_{1N1},\ldots,\rho^*_{1Nt-1}=0$
\item[(iii)] If $\rho^*_{0Nt}=\text{INR}_\text{c}$ then $\rho^*_{0N1},\ldots,\rho^*_{0Nt-1}=\text{INR}_\text{c}$
\item[(iv)] If $\rho^*_{1Nt}=\text{INR}_\text{c}$ then $\rho^*_{1Nt+1},\ldots,\rho^*_{1NT}=\text{INR}_\text{c}$
\end{enumerate}
\label{thm:noisyProfile}
\end{theorem}
 
\begin{proof}
See Appendix~\ref{proof:noisyProfile}.
\end{proof}

Due to the generality of our problem formulation in~(\ref{eqn:nsenseOptProb}), the exact form of the power distribution cannot be derived in closed form. However, Theorem~\ref{thm:noisyProfile} proves a very important result about the monotonicity of power allocation across time-slots. The optimal power profile is non-increasing in time if the start state is $\hat{s}_0=0$. That is the secondary gets \emph{paranoid} over time since it does not know when the primary transmitter will start transmitting. So it is better for the secondary to send more power in the initial time-slots and become more conservative as time progresses. 
In contrast, if $\hat{s}_0=1$, then the secondary bets more power in the later time-slots as there is a higher chance that the primary will turn off in those slots, thereby creating a better channel for the secondary flow.  
In the next section, we show that the above result can be significantly strengthened for the special case of perfect state estimate $\hat{s}_0=s_0$. 

\subsection{Special Case I: Secondary with perfect estimate of primary state, $\hat{s}_0 = s_0$\label{sec:psense}}

In this section we assume that there is no error in secondary's estimate of $s_{0}$. When the starting state is perfectly known, each of the $T$ subchannels shown in Figure~\ref{fig:multiplexedCW} behave as parallel channels. 
For such a channel capacity can be achieved by sending at a constant power with receiver side channel side information \cite{gv97}. We show below that Gaussian codewords with power levels $\rho_{st}=\rho_{sSt}+\rho_{sNt}$ achieves the capacity if the sensing is error-free.

\begin{theorem} [Capacity with perfect sensing, $\hat{s}_{0}=s_{0}$]
With perfect sensing, the capacity for the discrete cognitive interference channel is given by,
\begin{equation*}
C= \sum\limits_{s_0\in S} \frac{\pi(s_0)}{T+1}  \sum\limits_{t=1}^{T} \max_{p(X_2^{(t)}|S_0)} \sum\limits_{s\in S}  p(s|s_0) I(X_2^{(t)}; Y_2^{(t)}|s_0,s).
\end{equation*}
\label{thm:mainThm}
\end{theorem}
\begin{proof}
See Appendix~\ref{proof:mainThm}. 
\end{proof}

The above result for finite input-output alphabets extends to continuous alphabets such that $I(X_2^{(t)}; Y_2^{(t)}|s_0,s)=\beta_s(t)\mathcal{C}_{s0} +\bar{\beta}_s(t)\mathcal{C}_{s1}$, $\sum\limits_{s_0\in S} \sum\limits_{t=1}^{T} \pi(s_0) \rho_{st}\leq \text{SNR}_2$, $\rho_{st} \leq \frac{\text{INR}_\text{gap}}{|h_{21}|^2} \text{ for } t = 1,\cdots,T$ and $p(X_2^{(t)}|s_0)\sim \mathcal{N}(0,\rho_{st})$. 
In order to find the power allocation $\rho_{st}$, we have to solve an optimization problem similar to (\ref{eqn:genieCon}).
When sensing is perfect $\hat{s}_{0}=s_{0}$ and $P(\hat{s}_{i}|s_{0})=1$. There is no missed detection or false alarm, i.e. $\mathcal{P}_{M}=0$ and $\mathcal{P}_{F}=0$. So, the rate and all the constraint equations  for the perfect sensing protocol can be obtained by making these substitutions in~(\ref{eqn:nsenseOptProb}).

\begin{theorem}[Optimal Layer~$2$ power]
For the optimization problem given in (\ref{eqn:nsenseOptProb}) with $\mathcal{P}_{M}=\mathcal{P}_{F}=0$ and $\hat{s}_{0}=s_{0}$, the optimal power of the Layer~$2$ Gaussian codeword is given by, 
$\rho_{sSt}^*=\left(\min \left(\text{SIC}_\text{C}, \text{INR}_\text{C}, \text{SNR}_2 \right)\right)^+,$
where $\text{SIC}_\text{C}, \text{INR}_\text{C}$ are defined in Theorem~\ref{thm:genie} and $x^+=\max(x,0)$.
\label{thm:PSol}
\end{theorem}
\begin{proof}
See Appendix~\ref{proof:PSol}.
\end{proof}
Due to the perfect estimate of the primary starting state, the rate splitting done at secondary is optimal, which allows us to find the exact value of the Layer 2 codewords. The monotonicity properties of the power levels still hold for the perfect sensing case. 
\begin{theorem}[Monotonic Layer~$1$ profile]
For the optimization problem given in (\ref{eqn:nsenseOptProb}) with $\mathcal{P}_{M}=\mathcal{P}_{F}=0$, 
$\rho^*_{0N1} \geq \rho^*_{0N2} \geq \ldots \geq \rho^*_{0NT} \text{ and } \rho^*_{1N1} \leq \rho^*_{1N2} \leq \ldots \leq \rho^*_{1NT}.$ Additionally, 
\begin{enumerate}
\item[(i)] If $\rho^*_{0Nt}=0$ then $\rho^*_{0Nt+1},\ldots,\rho^*_{0NT}=0$
\item[(ii)] If $\rho^*_{1Nt}=0$ then $\rho^*_{1N1},\ldots,\rho^*_{1Nt-1}=0$
\item[(iii)] If $\rho^*_{0Nt}=\text{INR}_\text{c}$ then $\rho^*_{0N1},\ldots,\rho^*_{0Nt-1}=\text{INR}_\text{c}$
\item[(iv)] If $\rho^*_{1Nt}=\text{INR}_\text{c}$ then $\rho^*_{1Nt+1},\ldots,\rho^*_{1NT}=\text{INR}_\text{c}$
\end{enumerate}
\label{thm:PSprofile}
\end{theorem}

\begin{proof}
The proof follows directly from Theorem~\ref{thm:noisyProfile} for $\hat{s}_{0}=s_{0}$ and $\mathcal{P}_{M}=\mathcal{P}_{F}=0$.
\end{proof}
The intuition for the above results is as follows. It is always better to allocate more power to Layer~$2$ codewords (while satisfying the power and interference constraints) as it has a higher contribution towards the secondary rate $R_{2}$. For a fixed $|h_{12}|$, below a certain $\text{INR}_\text{gap}$, this upper bound is zero and all the power goes to the Layer~$1$ codeword. As $\text{INR}_\text{gap}$ increases, the proportion of the Layer~$2$ codeword keeps increasing and in the end all the power is put into the Layer~$2$ codeword. In short, we have to do layered water-filling for each time slot in the block as shown in Figure~\ref{fig:layered-water-filling}.

Next, we show that the opportunistic superposition of \cite{py07} is a special case of the coginitive protocol when no sensing is done. 
\subsection{Special Case II: Secondary with no information about primary state $s_0$ \label{sec:no info}}
Consider the special case when the secondary does not sense the channel at all, i.e. the secondary transmitter only knows the statistics of the primary traffic $\pi(s_{0})$ and $f_{T}(\tau)$. Alternately, the estimate $\hat{s}_0$ is so noisy that it does not provide any information about $s_0$. A similar analysis of this special case can also be found in \cite{ru96}.
Out of the available power $\text{SNR}_2$, $\rho_{S}=\alpha \text{SNR}_2$ is assigned to the Layer~$2$ codeword and $\rho_{N}=(1-\alpha) \text{SNR}_2$ is assigned to the Layer~$1$ codeword. 
The average power constraint for the secondary is given by $\rho_{N}+\rho_{S} \leq \text{SNR}_{2}$ and the INR constraint can be written as $\rho_{N}+\rho_{S}  \leq  \frac{\text{INR}_\text{gap}}{|h_{21}|^2} $.  
After decoding the single user codeword, the residual capacity of the channel is given by $\mathcal{C}\left(\frac{\text{INR}_2}{1+\alpha \text{SNR}_2}\right)$. 
To ensure primary is decodability the SIC constraint is given by, 
$R_1 \leq \mathcal{C}\left(\frac{\text{INR}_2}{1+\alpha \text{SNR}_2}\right)$ 
$\left(\text{i.e. }  \alpha \leq \frac{|h_{12}|^2(\text{INR}_\text{gap}+1)-1}{\text{SNR}_2}\right)$. 
For such an $\alpha$, a secondary rate of $R_2(\alpha)=\mathcal{C}\left( \frac{(1-\alpha)\text{SNR}_2}{1+\text{INR}_2+\alpha \text{SNR}_2} \right)+\mathcal{C}\left( \alpha \text{SNR}_2 \right)$ can be achieved. 
Gaussian codebooks are optimal in this case too by Lemma~\ref{lem:Gopt}. 
The optimal value of the superposition fraction ($\alpha^{*}$) is one which maximizes $R_2(\alpha)$ while satisfying all the constraints. 
\begin{theorem}[No-sensing Capacity]
For a cognitive interference channel where the secondary does not sense the channel, the optimal superposition fraction is given by 
\begin{equation*}
\alpha^* = \left\{ \begin{array}{ll} 
0 & \mbox{if $\text{INR}_2 \leq \frac{\text{SNR}_1}{1+\text{INR}_\text{gap}}$}\\ 
\frac{\text{SIC}_\text{C}}{\text{SNR}_2}& \mbox{$\frac{\text{SNR}_1}{1+\text{INR}_\text{gap}} < \text{INR}_2 < \frac{\text{SNR}_1(1+\text{SNR}_2)}{1+\text{INR}_\text{gap}} $} \\ 
1 & \mbox{if $\text{INR}_2 \geq \frac{\text{SNR}_1}{1+\text{INR}_\text{gap}}(1+\text{SNR}_2)$}
\end{array},
\right.
\end{equation*}
and the capacity is given by $R_{2}(\alpha^{*})$.
\label{thm:nonBCap}
\end{theorem}

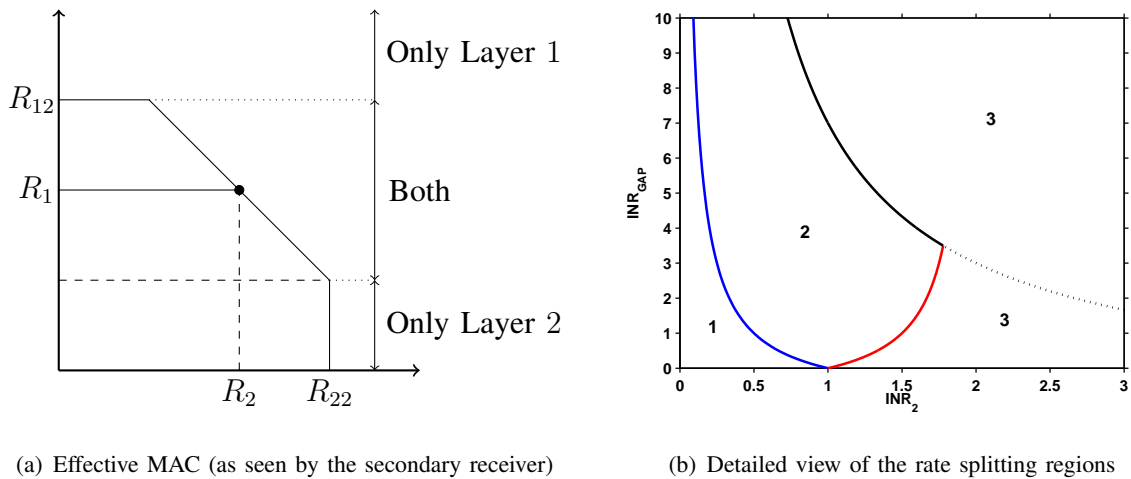
\begin{figure}[htbp]
\centering
\subfigure[Effective MAC (as seen by the secondary receiver)]{
	\centering
	\tikz[scale=1.2]{
		\draw [->,thick] (1,1) -- (5,1);
		\draw [->,thick] (1,1) -- (1,5);
		\draw (1,4) -- (2,4);
		\draw (4,1) -- (4,2);
		\draw (2,4) -- (4,2);
		\draw (1,3) -- (3,3);
		\draw [fill=black] (3,3) circle (.05cm);
		\draw [dashed] (3,1) -- (3,3);
		\draw [dashed] (1,2) -- (4,2);
		\draw (.7,4) node {$R_{12}$};
		\draw (.75,3) node {$R_{1}$};
		\draw (3,.75) node {$R_{2}$};
		\draw (4,.75) node {$R_{22}$};
		\draw [<->] (4.5,1) -- (4.5,2);
		\draw [<->] (4.5,2) -- (4.5,4);
		\draw [->] (4.5,4) -- (4.5,5);
		\draw [dotted] (4,2) -- (4.5,2);
		\draw [dotted] (2,4) -- (4.5,4);
		\draw (5.6,1.5) node {Only Layer~$2$};
		\draw (5.6,4.5) node {Only Layer~$1$};
		\draw (5,3) node {Both};
	}
	\label{fig:N1geqN2A}
}
\subfigure[Detailed view of the rate splitting regions]{
	\centering
	\includegraphics[width=3in]{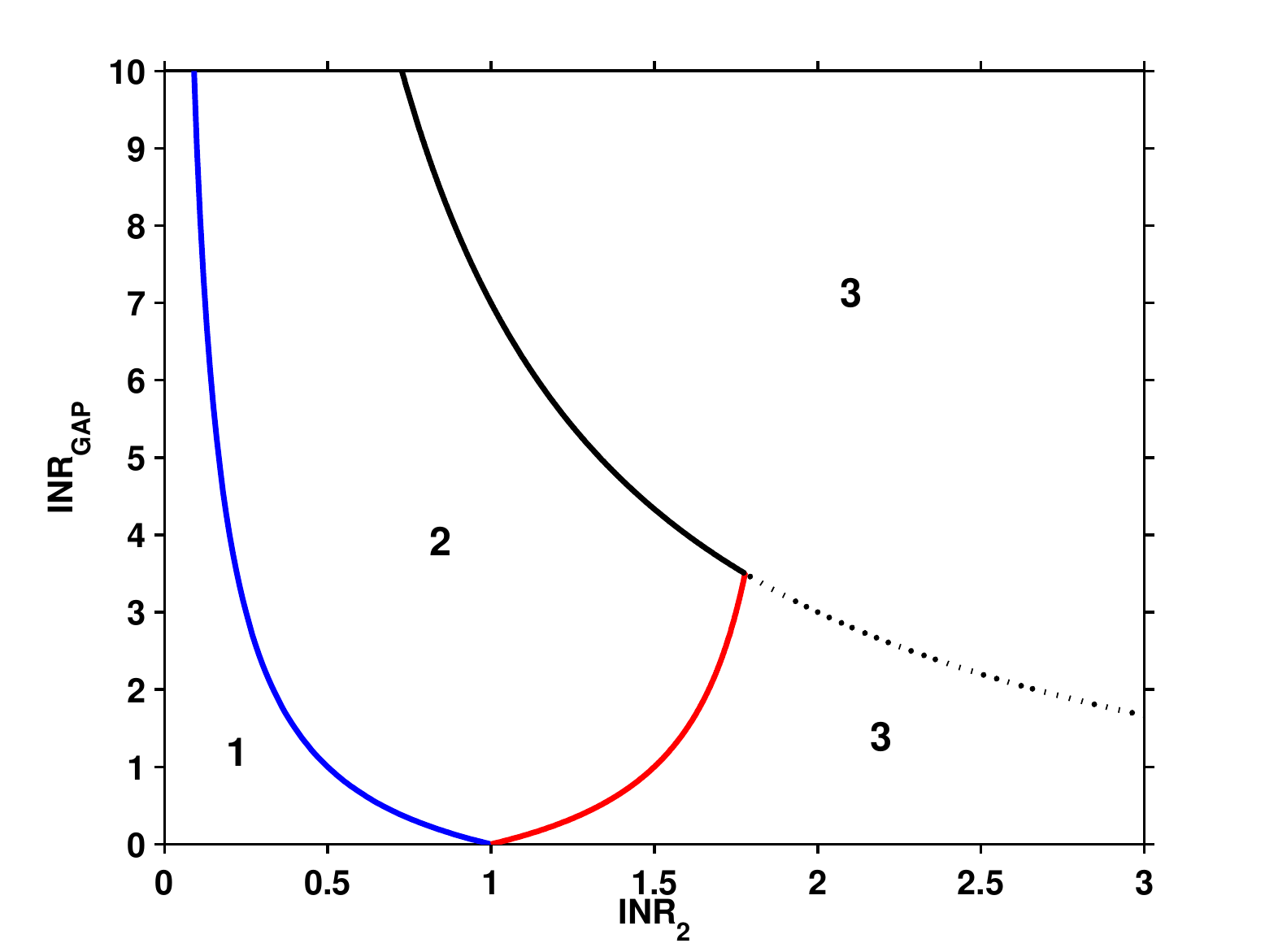}
	\label{fig:N1geqN2B}
}
\caption{(A) The effective MAC as seen by the secondary receiver, where $R_{1}=\mathcal{C}(\frac{\text{INR}_{2}}{1+ \alpha \text{SNR}_{2}})$ is the decodable primary rate as seen by the secondary receiver, $R_{12}=\mathcal{C}(\text{INR}_{2})$ is the rate supported by the cross channel between the primary transmitter and the secondary receiver, $R_{2}(\alpha)$ is the rate achievable by the secondary and $R_{22}=\mathcal{C}(\text{SNR}_{2})$ is the maximum rate achievable by the secondary if there is no primary. (B) Detailed view of the regions of operation for the secondary user for $\text{SNR}_1=\text{SNR}_2=7$ as a function of the two variables $\text{INR}_{\text{gap}}$ and $\text{INR}_{2}$. The secondary user uses Layer~$1$ codewords in Region~$1$, Layer~$2$ codewords in Region~$3$ and a superposition code with both layers in Region~$2$. Note that $\text{INR}_{\text{gap}}=0$ is not a part of any region.
\label{fig:N1geqN2}}
\end{figure}

\begin{proof}
See Appendix~\ref{proof:nonB}. 
\end{proof}
From the secondary receiver's perspective, the equivalent channel is a MAC. The fixed primary converts the equivalent MAC rate region (pentagon) to a single line (Figure~\ref{fig:N1geqN2A}). Even though the rate splitting assumes sequential decoding, it turns out to be optimal \cite{cov91} because the two code layers at the secondary, makes the rate tuple a corner point of a three user virtual MAC \cite{ru96} consisting of the two code layers of secondary and the primary. 

\begin{remark}[Persistent Primary\label{sec:nonBursty}]
If the primary has persistent data ($\pi(1)=1$), the rate that the secondary can achieve is the same as proved above, i.e. the effective channel in the no-sensing case is a compound channel and the secondary has to code for the worst case. 
\end{remark}

The optimal distribution puts as much available power in $\rho_{1S}$ as possible without violating the SIC, INR and power constraints. This can be thought of a layered water-filling as shown in Figure~\ref{fig:layered-water-filling}. The layer due to the primary codeword is fixed. Available power is first assigned to the Layer 2 codeword. When the fixed layer is very high so that no power is left to put in the Layer 1 codeword, $\alpha^*=0$ and if the fixed layer touches the noise floor, all the power is put in the Layer 1 codeword, $\alpha^*=1$.
The idea of opportunistically doing interference cancellation has also been analyzed in \cite{zc09a,py07} and is a restatement of the rate splitting approach introduced in \cite{ru96} for achieving time-sharing without cooperating encoders. The difference from \cite{py07} is the effect of the secondary transmissions on the primary which gives rise to the power profile and the proof of optimality using rate splitting and sequential decoding.

Figure~\ref{fig:N1geqN2B}  shows a detailed view of the the different rate splitting regions as a function of $\text{INR}_{\text{gap}}$ and $\text{INR}_{2}$,  for fixed $\text{SNR}_1$, $\text{SNR}_2$. In Region~1, the secondary user treats the primary data as noise ($\alpha^*=0$). In region~2 the secondary uses both layers. In Region~3 the secondary receiver first decodes the primary's data and then decodes its own ($\alpha^*=1$).
The Layer~$1$ protocol region is always the same w.r.t. $\text{INR}_1$. The increase in the Layer~$2$ protocol region with the increase in the $|h_{21}|^2$ can be attributed to the interference constraint at the primary receiver. This increase comes at the expense of the superposition protocol region. 
When $|h_{21}|^2$ increases, the power available to the secondary user keeps on decreasing due to the interference constraint. This decreases the power left for the Layer~$1$ codeword. Hence at higher  $|h_{21}|^2$, a part of the mixed protocol region gets converted to a multiuser protocol region as there is no more power left to put into the single user codeword.

\begin{figure}[ht]
	\centering
	\includegraphics[width=3in]{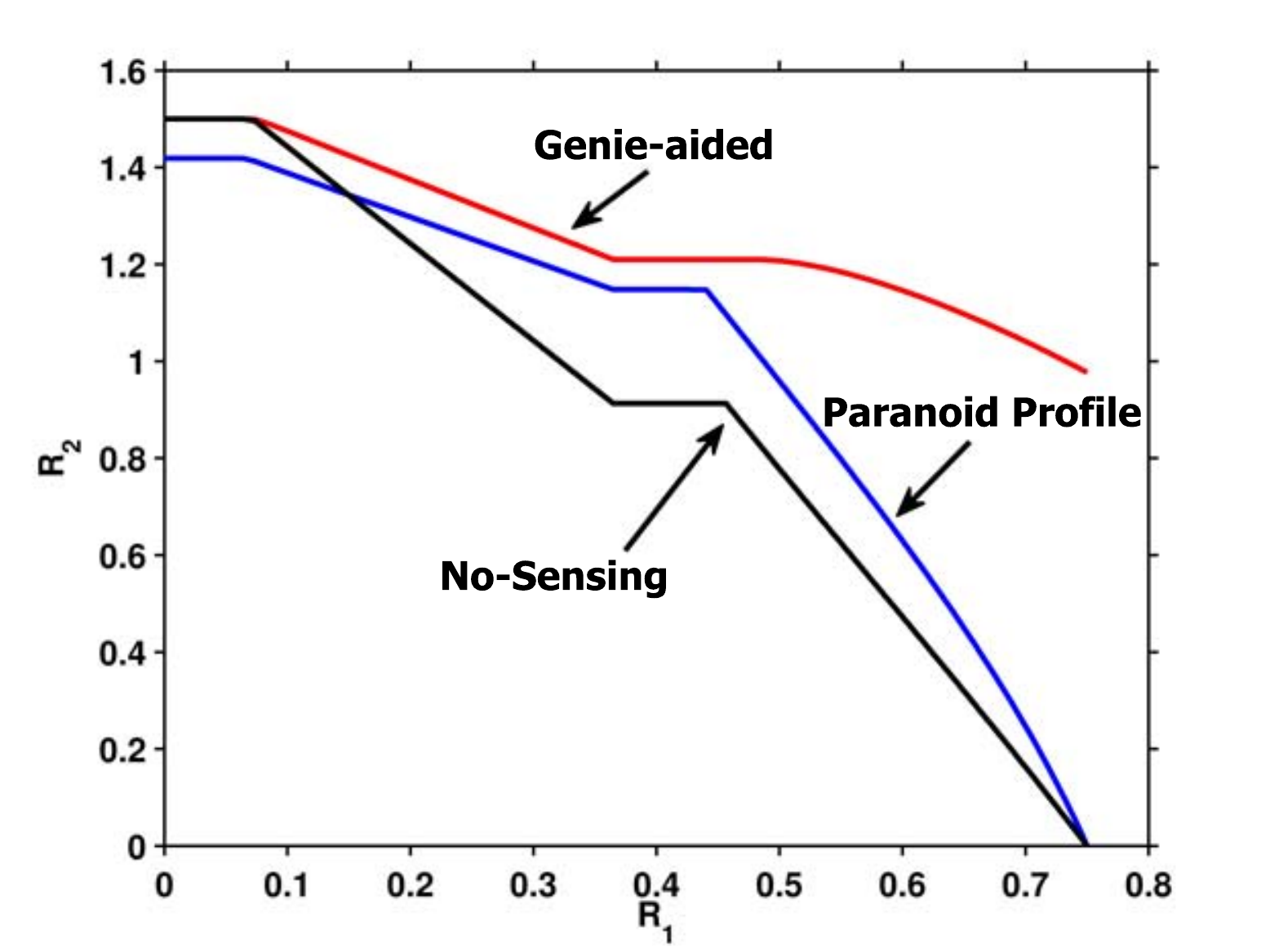} 
	\caption{The no-information lower bound can outperform perfect sensing scheme with a finite sensing overhead depending on the $\text{INR}_{\text{gap}}$, the blocklength $T$. Here, $P_1=P_2=7, \beta=0.5, \pi(0)=\pi(1)=0.5, |h_{12}|^2=|h_{21}|^2=0.5$ and $T=10$.
	\label{fig:rateall}}
\end{figure}

\begin{figure}[ht]
	\centering
	\includegraphics[width=3in]{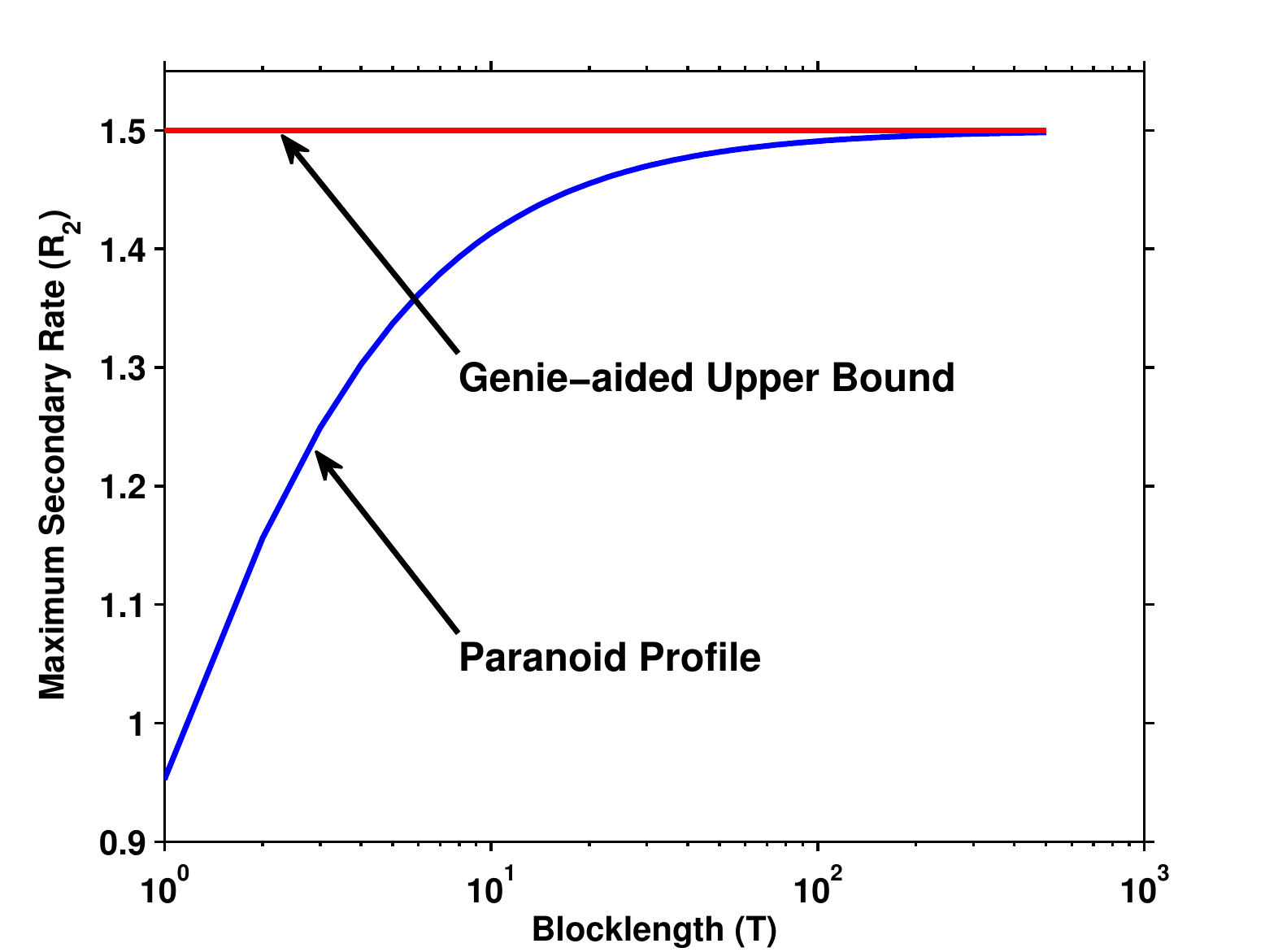}
	\caption{The maximum $R_2$ achieved (two extreme rate points on the y-axis in Figure~\ref{fig:rateall} where $R_{1}=0$ ) by the paranoid profile scheme gets closer to the genie-aided upper bound as $T$ increases which is due to smaller sensing overhead. Here, $P_1=P_2=7, \beta=0.5, \pi(0)=\pi(1)=0.5, |h_{12}|^2=|h_{21}|^2=0.5$.
	\label{fig:maxR2}}
\end{figure}

\begin{figure}[ht]
	\centering
	\includegraphics[width=3in]{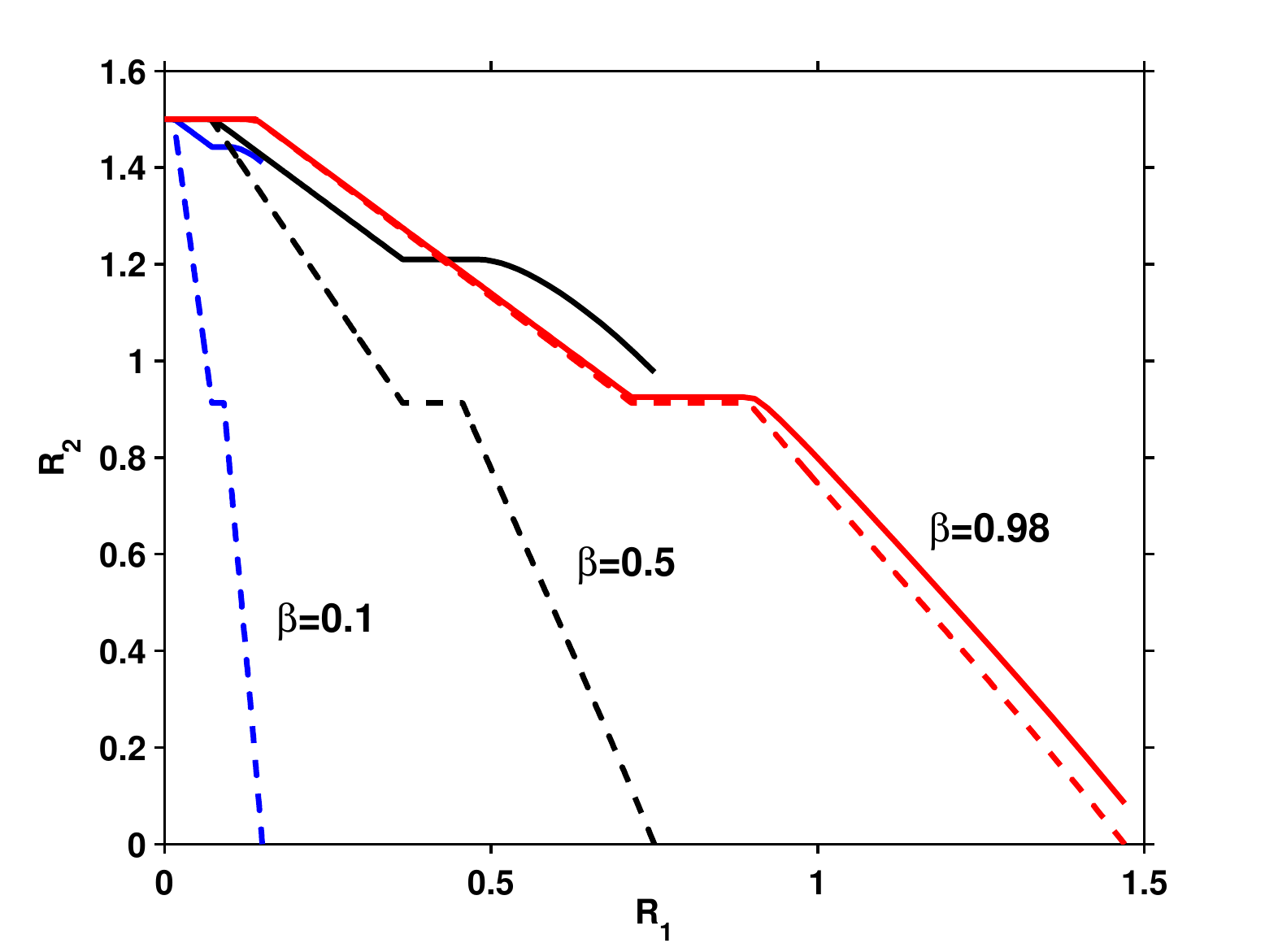}
	\caption{Comparison of no-sensing scheme and the Genie-aided scheme for different persistence ($\beta$) of the primary user's data and $|h_{12}|^2=|h_{21}|^2=0.5$. The dashed lines are for the no-sensing scheme and the solid lines are for the genie-aided scheme. 
	\label{fig:bndBurstyP}}
\end{figure}

The paranoid power profile is optimal if the sensing is perfect. However, even if the sensing is noise free, the time spent in sensing is still an overhead. This overhead can be high enough for the no-sensing scheme to outperform the perfect sensing scheme in some regimes, as shown in Figure~\ref{fig:rateall}.
Along the curve, points close to the right side  ($R_2=0$) are obtained for $\text{INR}_\text{gap}=0$ and the points to the left side ($R_1=0$) are obtained for $\text{INR}_\text{gap}=\infty$. 
Figure~\ref{fig:maxR2} plots the two extreme rate points on the y-axis in Figure~\ref{fig:rateall} (where $R_{1}=0$) for different block sizes. The above-mentioned loss due to sensing, decreases as $T$ increases.
Finally, the bounds are closer together when the residual capacity of the primary channel is smaller as shown in Figure~\ref{fig:bndBurstyP}. Different values of $\beta$ has no effect on the no-information lower bound. 
But the genie-aided upper bound is able to get a higher rate if the channel is idle for a longer period of time. This means, when $\beta$ is high, the performance of both these bounds are close.


\section{Conclusion}
We approximate an interference channel in the presence of a fixed sporadic primary flow as a block activity channel. Such a block activity model can be broken into parallel channels in time and multiplexed codebooks can be used when there is perfect sensing. We derived a paranoid scheme for the secondary user when the primary user is transmitting below capacity and show that rate splitting at the secondary transmitter with sequential decoding is optimal when the estimation of the primary starting state is noise-free. Depending on the starting state of the primary during a block, the optimal power profile for Gaussian inputs is either growing or decaying in power as a function of the time slot. If the sensing overhead is considered, we showed that the paranoid scheme approaches the genie-aided scheme for large block lengths.
Finally we show numerically that the paranoid and genie aided schemes approach the no-information scheme when the primary channel is operating close to its capacity.
\appendices


\section{\label{proof:genie}}
Proof of Theorem~\ref{thm:genie}: 
We derive a special case of the water-filling result \cite{gal68} for a channel with two states, which will be used to prove the optimal power allocation for the genie-aided case.  

\begin{lemma}
For the optimization problem given by
\begin{eqnarray}
&\max\limits_{\rho_0,\rho_1\geq 0} &\bar{\beta} \mathcal{C}(\rho_0)+\beta\mathcal{C}(\frac{\rho_{1}}{1+\alpha}) \nonumber\\
&\text{subject to }&\bar{\beta} \rho_0+\beta\rho_1 \leq \gamma \text{ and } \rho_1 \leq \delta 
\label{eqn:lemOptProb}
\end{eqnarray}
with non-negative parameters $(\alpha, \beta,\gamma, \delta)$, the maximizing $\rho_0, \rho_1$ is characterized as follows.

\begin{equation*}
(\rho_0,\rho_1) = \begin{cases} 
(\frac{\gamma}{\bar{\beta}}, 0) & \mbox{if $\rho^*_1 < 0 \leq \delta$} \hspace{1cm} (A)\\ 
(\rho_0^*, \rho_1^*) & \mbox{if $0 \leq \rho^*_1 \leq \delta$} \hspace{1cm} (B) \\ 
(\frac{\gamma-\beta\delta}{\bar{\beta}}, \delta) & \mbox{if $0\leq \delta < \rho^*_1$} \hspace{1cm} (C) 	
\end{cases},
\end{equation*}
where $\rho_0^*=\gamma+\beta\alpha, \rho_1^*=\gamma-\bar{\beta}\alpha$  and $\bar{\beta}=1-\beta$.
\label{lem:optProb}
\end{lemma}

\begin{proof}
Let us start with assuming that there is no constraint on $\rho_1$. 
The optimal solution is given by $\rho_0 = \left( \frac{1}{\lambda} - 1 \right)^+, \rho_1 = \left( \frac{1}{\lambda} - 1-\alpha \right)^+$,
where $\lambda$ is chosen such that $\bar{\beta} \rho_0+\beta\rho_1 = \gamma$.
If $\frac{1}{\lambda} > 1 + \alpha$, then $\frac{1}{\lambda} = \gamma + \bar{\beta} + \beta (1+ \alpha)$.
Therefore, $\rho_0^*=\gamma+\beta\alpha$ and $\rho_1^*=\gamma-\bar{\beta}\alpha$. Note $\frac{1}{\lambda} > 1 + \alpha$ also ensures $\rho_1^* \geq 0 $ and by assumption $\rho_1^* \leq \delta$. This proves part B of the Lemma.

If $\frac{1}{\lambda} \leq 1 + \alpha$, $\rho_1=0$ and $\lambda$ is such that $\bar{\beta}\rho_0=\gamma$, then $\rho_0=\frac{\gamma}{\bar{\beta}}, \rho_1 = 0$. This proves part A of the solution.  
As long as $\rho_1^*$ lies below $\delta$ (i.e. the extra constraint on $\rho_1$ is not active at the solution), the same solutions hold. If $\rho_1^* > \delta$, $\rho_1=\delta$, $\rho_0=\frac{\gamma-\beta\delta}{\bar{\beta}}$. 
\end{proof}

\begin{figure}[ht]
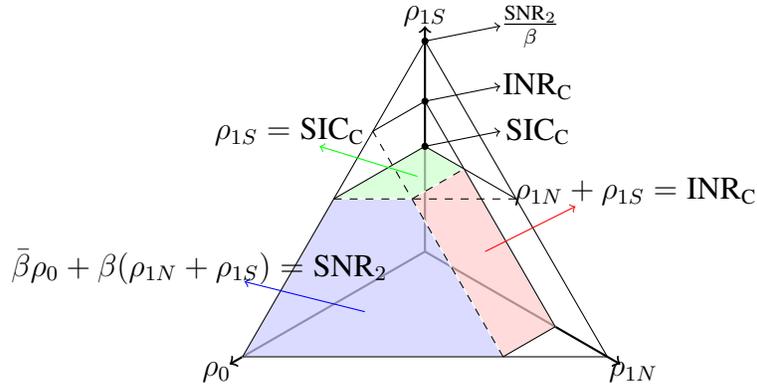

\centering
	\centering
	\tikz[scale=0.2]{
		\draw [->, thick] (0,0) -- (0,15);
		\draw [->, thick] (0,0) -- (15*.866,-15*.5);
		\draw [->, thick] (0,0) -- (-15*.866,-15*.5);
		\node at (16*.866,-16*.5) {$\rho_{1N}$};
		\node at (0,15+.7) {$\rho_{1S}$};
		\node at (-16*.866,-16*.5) {$\rho_0$};
		\filldraw [fill=green!20, draw=white, fill opacity=0.7] (0,7) -- (-7*.866,-7*.5+7) -- (-1*.866,-7*.5+7) -- (3*.866,-3*.5+7) -- (0,7);
		\filldraw [fill=red!20, draw=white, fill opacity=0.7]  (-1*.866,-7*.5+7) -- (3*.866,-3*.5+7) -- (10*.866,-10*.5) -- (6*.866,-14*.5) -- (-1*.866,-7*.5+7);
		\filldraw [fill=blue!20, draw=white, fill opacity=0.7] (-7*.866,-7*.5+7) -- (-1*.866,-7*.5+7) -- (6*.866,-14*.5) -- (-14*.866,-14*.5) -- (-7*.866,-7*.5+7);
		\draw (0,14) -- (14*.866,-14*.5);
		\draw (0,14) -- (-14*.866,-14*.5);
		\draw (-14*.866,-14*.5) -- (14*.866,-14*.5);
		\draw (0,10) -- (10*.866,-10*.5);
		\draw (10*.866,-10*.5) -- (6*.866,-14*.5);
		\draw (0,10) -- (-4*.866,-4*.5+10);
		\draw [dashed] (6*.866,-14*.5) -- (-4*.866,-4*.5+10);
		\draw (0,7) -- (-7*.866,-7*.5+7);
		\draw (0,7) -- (7*.866,-7*.5+7);
		\draw [dashed] (-7*.866,-7*.5+7) -- (7*.866,-7*.5+7);
		\draw [dashed] (-1*.866,-7*.5+7) -- (3*.866,-3*.5+7);
		\draw [->, color=red] (4,0) -- (10,3);
		\node at (14,4) {$\rho_{1N} + \rho_{1S} = \text{INR}_\text{C}$};
		\draw [->, color=green] (-.5,5) -- (-7,7);
		\node at (-9,8) { $\rho_{1S} = \text{SIC}_\text{C}$};
		\draw [->, color=blue] (-4,-4) -- (-12,-2);
		\node at (-15,-1) {$\bar{\beta} \rho_0 + \beta (\rho_{1N} + \rho_{1S}) = \text{SNR}_2$};
		\draw [fill=black] (0,14) circle (.2cm);
		\draw [->] (0,14) -- (5,15);
		\node at (7, 15) {$\frac{\text{SNR}_2}{\beta}$};
		\draw [fill=black] (0,10) circle (.2cm);
		\draw [->] (0,10) -- (5,11);
		\node at (7.5, 11) {$\text{INR}_\text{C}$};
		\draw [fill=black] (0,7) circle (.2cm);
		\draw [->] (0,7) -- (5,8);
		\node at (7.5, 8) {$\text{SIC}_\text{C}$};
	}
	\caption{General Constraint set for the perfect sensing optimization problem.
	\label{fig:3dConSet}}
\end{figure}

The constraint set defined by the average power constraint at the secondary transmitter, the INR constraint at the primary receiver, and the SIC constraint at the secondary receiver is schematically shown by the three colored planes in Figure~\ref{fig:3dConSet}.

First consider the case when the SIC constraint is not active i.e. when $\text{SIC}_{\text{C}}\geq \min\left( \text{SNR}_{2},\text{INR}_{\text{C}}\right)$.
In this case, the solution lies on the $\rho_0-\rho_{1S}$ plane of Figure~\ref{fig:3dConSet}. This is because $R_2(\rho_0, \rho_{1N}, \rho_{1S}) \leq R_2(\rho_0, \rho_{1N}-\kappa, \rho_{1S}+\kappa)$ as long as $\kappa \leq \rho_{1N}$, i.e. we can achieve a higher rate by taking away some power from $\rho_{1N}$ and giving it to $\rho_{1S}$ as long as all the constraints are satisfied. Therefore, to maximize the rate, all the available power is allocated to $\rho_{1S}$ since there is no SIC constraint.
When we put $\rho_{1N}=0$, the optimization problem looks like Equation~(\ref{eqn:lemOptProb}) with $\alpha=0, \gamma=\text{SNR}_{2}$ and $\delta=\text{INR}_{\text{C}}$. By using Lemma~\ref{lem:optProb}, the solution is given by Equation~\ref{eqn:genie1}.

Finally, if $\text{SIC}_\text{C} < \min \left( \text{SNR}_2, \text{INR}_\text{C} \right)$, the SIC constraint is active at the solution and the solution lies on the SIC constraint plane of Figure~\ref{fig:3dConSet}. In contrast to the above cases, we cannot put all the power from $\rho_{1N}$ in $\rho_{1S}$. We increase $\rho_{1S}$ till $\rho_{1S}=\text{SIC}_\text{C}$ which equals the SIC constraint. Now the problem is same as Equation~\ref{eqn:lemOptProb} with $\alpha=|h_{12}|^2(1+\text{INR}_\text{gap}+\text{SNR}_1)-1, \gamma=\text{SNR}_2-\beta\text{SIC}_\text{C} \text{ and } \delta=\text{INR}_\text{C}-\text{SIC}_\text{C}$. Using Lemma~\ref{lem:optProb}, and the solution is given by Equation~\ref{eqn:genie2}.


\section{\label{summChange}}
\begin{lemma}[Change in Order of Summation]
Given a probability mass function, $f(t), 1\leq t \leq T$ of a discrete random variable, and positive real numbers $(\rho_1,\ldots,\rho_T), \alpha$ and $\beta$,
\begin{equation*}
\sum\limits_{j=1}^{T+1} f(j) \left[ \sum\limits_{i=1}^{j-1}C\left( \frac{\rho_i}{\alpha} \right) + \sum\limits_{i=j}^{T}C\left( \frac{\rho_i}{\beta} \right)\right]
= \sum\limits_{i=1}^{T} \left[ (1-F(t))C\left( \frac{\rho_i}{\alpha} \right) + F(t)C\left( \frac{\rho_i}{\beta} \right),  \right]
\end{equation*}
where $C(x)=\log(1+x)$,  $F(j)=\sum_{i=1}^{j}f(i), 1\leq j \leq T$ and $F(0)=0, F(T+1)=1$.
\label{lem:chOrder}
\end{lemma}

\begin{proof}
Let us first consider the first term of the left hand side with $\alpha=1$,
\begin{align*}
\sum\limits_{j=1}^{T+1} f(j) \sum\limits_{i=1}^{t-1} c(\rho_i)
=& f(2)c(\rho_1)+ f(3) [ c(\rho_1) +c(\rho_2)]+ 
\ldots + f(T+1)[c(\rho_1)+\ldots+c(\rho_T)]\\
=& c(\rho_1)[f(2)+\ldots+f(T+1)] + \ldots + c(\rho_T) f(T+1)= \sum\limits_{i=1}^{T} c(\rho_i) \sum\limits_{j=i+1}^{T+1} f(j).
\end{align*}
From the definition of the probability distribution function, $1-F(j)= \sum\limits_{i=1}^{T+1}f(i) - \sum\limits_{i=1}^{j}f(i) = \sum\limits_{i=j+1}^{T+1}f(i)$.
Hence, $\sum\limits_{j=1}^{T+1} f(j) \sum\limits_{i=1}^{t-1} c(\rho_i)=\sum\limits_{i=1}^{T} c(\rho_i) \sum\limits_{j=i+1}^{T+1} f(j)=\sum\limits_{i=1}^{T}(1-F(t))c(\rho_i)$. \hspace{.1cm}Similarly,  $\sum\limits_{j=1}^{T+1} f(j) \sum\limits_{i=t}^{T} c(\rho_i)=\sum\limits_{i=1}^{T}F(t)c(\rho_i),$ which completes the proof.
\end{proof}

\section{\label{proof:noisyProfile}}
Proof of Theorem~\ref{thm:noisyProfile}: We will drop the subscript $0$ on $s,\hat{s}$ to reduce clutter. The Lagrangian for the optimization problem in Theorem~\ref{thm:noisyProfile} can be written as,
\begin{equation*}
L(\bm{\rho},\lambda_1,\lambda_2)=R_{2}(\bm{\rho})-\lambda_1g_1(\bm{\rho})-\sum\limits_{s,t}\lambda_{2\hat{s}t}g_{2\hat{s}t}(\bm{\rho})-\sum\limits_{s,t}\lambda_{3\hat{s}t}g_{3\hat{s}t}(\bm{\rho})-\sum\limits_{s,t}\lambda_{4\hat{s}t}\rho_{\hat{s}Nt}-\sum\limits_{s,t}\lambda_{5\hat{s}t}g_{\hat{s}St}.
\end{equation*}
The partial derivatives of the Lagrangian is given by,
\begin{align*}
\frac{\delta L}{\delta \rho_{\hat{s}Nt}} 
= &    -\sum_s \pi(s) p(\hat{s}|s) \left( \frac{\bar{\beta}_s(t)}{1+\rho_{\hat{s}Nt}+\rho_{\hat{s}St}} + \frac{\beta_s(t)}{1+\text{INR}_2+\rho_{\hat{s}Nt}+\rho_{\hat{s}St}} \right) -\lambda_1 \sum_s \pi(s) p(\hat{s}|s) - \lambda_{3\hat{s}t} -\lambda_{4\hat{s}t}\\
\frac{\delta L}{\delta \rho_{\hat{s}St}} 
= &    -\sum_s \pi(s) p(\hat{s}|s) \left( \frac{\bar{\beta}_s(t)}{1+\rho_{\hat{s}Nt}+\rho_{\hat{s}St}} + \frac{\beta_s(t)}{(1+\text{INR}_2+\rho_{\hat{s}Nt}+\rho_{\hat{s}St})(1+\text{INR}_2+\rho_{\hat{s}St})}\right) \\
 & - \lambda_1\sum_s \pi(s) p(\hat{s}|s)-\lambda_{2st}-\lambda_{3st} - \lambda_{5st}.
\end{align*}
The first order necessary conditions (the complementarity conditions are excluded for lack of space) are given by the following $12T+1$ equations,
$\frac{\delta L}{\delta \rho_{sNt}} =0, 
\frac{\delta L}{\delta \rho_{sSt}} =0, 
g_1(\rho) \geq 0, 
g_{2st} \geq 0, 
g_{3st} \geq 0, \text{and }
\rho_{sNt}, \rho_{sSt} \geq 0, \text{for }s\in \{0,1\}, t\in [1,T].$
If $|h_{12}|^2(\text{INR}_\text{gap}+1)<1$, then $\rho_{sSt}=0$. Hence $g_{2st}$, $\lambda_{2st}$ are redundant, and by complementarity $\lambda_{5st}=0$. Similarly, if $0<\rho_{sNt}<\text{INR}_\text{C}$ by complementarity $\lambda_{3st}=0$ and $\lambda_{4st}=0$. As the objective is an increasing function in each $\rho_{sNt}$, the average power constraint is met with equality, i.e. $\lambda_1 \geq 0$. 
For $|h_{12}|^2(1+\text{SNR}_\text{gap})<1$ and $0<\rho_{\hat{s}St}<\frac{P_0}{|h_{21}|^2}$, we have
$$\sum_s \pi(s) p(\hat{s}|s) \left( \frac{\bar{\beta}_s(t)}{1+\rho_{\hat{s}Nt}} + \frac{\beta_s(t)}{1+\text{INR}_2+\rho_{\hat{s}Nt}} \right) =\lambda_1 \sum_s \pi(s) p(\hat{s}|s) \text{ for all }s\in \{0,1\} \text{ and }t\in[1,T].$$
For a given $\hat{s}$ independent of any $t$, the right hand side of the above equation is constant. 
Rewriting the left hand side for $\hat{s}=0$,
\begin{align*}
\pi(0)(1-P_F)\left( \frac{\bar{\beta}_0(t)}{1+\rho_{0Nt}} + \frac{\beta_0(t)}{1+\text{INR}_2+\rho_{0Nt}} \right) + \pi(1)P_M\left( \frac{\bar{\beta}_1(t)}{1+\rho_{0Nt}} + \frac{\beta_1(t)}{1+\text{INR}_2+\rho_{0Nt}} \right) &= \text{const.}\\
\text{i.e. } \frac{\pi(0)(1-P_F)(1-F(t)) + \pi(1)P_MF(t)}{1+\rho_{0Nt}} + \frac{\pi(0)(1-P_F)F(t) + \pi(1)P_M(1-F(t))}{1+\text{INR}_2+\rho_{0Nt}} &=\text{const.}\\
\text{i.e. } \frac{G(t)}{1+\rho_{0Nt}} + \frac{H(t)}{1+\text{INR}_2+\rho_{0Nt}} &=\text{const.},
\end{align*}
where $G(t)=\pi(0)(1-P_F)(1-F(t)) + \pi(1)P_MF(t)$ and $H(t)=\pi(0)(1-P_F)F(t) + \pi(1)P_M(1-F(t))$. Rewriting the equation again,
\begin{equation*}
\begin{split}
\underbrace{\frac{G(t)}{1+\rho_{0Nt}} + \frac{H(t)}{1+\text{INR}_2+\rho_{0Nt}} }_{A_t} = \underbrace{\frac{G(t+1)}{1+\rho_{0Nt}} + \frac{H(t+1)}{1+\text{INR}_2+\rho_{0Nt}} }_{B_t} +~C_{t},
\end{split}
\end{equation*}
\begin{align*}
\text{where } C_t&=\frac{G(t)-G(t+1)}{1+\rho_{0Nt}} - \frac{H(t+1)-H(t)}{1+\text{INR}_2+\rho_{0Nt}}\\
      &=(\pi(0)(1-P_F)- \pi(1)P_M) (F(t+1)-F(t)) \left( \frac{1}{1+\rho_{0Nt}} - \frac{1}{1+\text{INR}_2+\rho_{0Nt}} \right).
\end{align*}
To make the right hand side $A_{t+1}$, we have to remove $C_t$ and replace $\rho_{0Nt}$ by $\rho_{0N(t+1)}$ in $B_t$. Since $C_t\geq 0$ when $\pi(0)(1-P_F)\geq \pi(1)P_M$, and $B_t$ is an decreasing function of $\rho_{0Nt}$, we have $\rho^*_{0Nt} \geq \rho^*_{0N(t+1)}$. Equality holds when $F_T(t)=F_T(t+1)$. As this is true for any $t \in [1, T]$, $\rho^*_{0N1} \geq \rho^*_{0N2} \geq \ldots \geq \rho^*_{0NT}$. 

 For a given $s$, say $s=0$, if there is a $t$ such that $\rho^*_{0Nt}=0$, then we will now prove by contradiction that $\rho^*_{0Nt+1}=\ldots=\rho^*_{0NT}=0$. Let us assume that $\rho^*_{0Nt+1}=\rho>0$. Define $R(y,z)=R_2(\rho_{0N1},\ldots,\rho_{0Nt}=y,\rho_{0Nt+1}=z,\ldots,\rho_{0NT},\rho_{1N1},\ldots,\rho_{1NT})$. Note $R_{2}(0,\rho)<R_{2}(\rho,0)$ for any $t \in [1,T]$ and any $\rho>0$ which satisfies all the constraints.
\begin{align*}
R_{2}(\rho,0)-R_{2}(0,\rho)=&(1-F(t))\mathcal{C}(\rho)+F(t)\mathcal{C}\left( \frac{\rho}{1+\text{INR}_2} \right)\\&-(1-F(t+1))\mathcal{C}(\rho)+F(t+1)\mathcal{C}\left( \frac{\rho}{1+\text{INR}_2} \right)\\
=& (F(t+1)-F(t))\left(\mathcal{C}(\rho) - \mathcal{C}\left( \frac{\rho}{1+\text{INR}_2} \right)\right)>0,
\end{align*}
since $\text{INR}_2>0$ and $F(t+1)\geq F(t)$. Hence if $\rho_{0Nt}=0$ then $\rho_{0Nt+1}=0$ and similarly for $t+2,\ldots,T$. We can prove the other parts of the theorem in the exact same way. For all previous time slots, the ordering follows the same order,
\begin{itemize}
\item If $\rho_{0Nt=0}$ then  $\rho^*_{0N1} \geq \rho^*_{0N2} \geq \ldots \geq \rho^*_{0Nt-1}>\rho^*_{0Nt}=\ldots=\rho^*_{0NT}=0$.

And due to the same reasons,
\item If $\rho_{1Nt}=0$ then  $\rho^*_{1N1}=\rho^*_{1N2}= \ldots = \rho^*_{1Nt-1}=0\leq \rho^*_{1Nt}\leq\ldots\leq \rho^*_{1NT}$.
\item If $\rho_{0Nt}=\frac{P_0}{|h_{21}|^2}$ then  $\rho^*_{0N1}=\rho^*_{0N2}= \ldots = \rho^*_{0Nt-1}=\frac{P_0}{|h_{21}|^2} \geq \rho^*_{0Nt}\geq\ldots\geq \rho^*_{0NT}$.
\item If $\rho_{1Nt}=\frac{P_0}{|h_{21}|^2}$ then  $\rho^*_{1N1} \geq \rho^*_{1N2} \geq \ldots \geq \rho^*_{1Nt-1}>\rho^*_{1Nt}=\ldots=\rho^*_{1NT}=\frac{P_0}{|h_{21}|^2} $.
\end{itemize}

 Note, for the case when $\text{SIC}_\text{C}>0$ then $\rho_{sSt}>0$. From Theorem~\ref{thm:PSol}, the value of $\rho_{sSt}$ is fixed. Hence, we can remove it from our problem statement along with $g_{2st}$ and $\lambda_{2st}$. The $\frac{\delta L}{\delta \rho_{sSt}} $ part of the Lagrangian follows in the same lines.

\section{\label{proof:mainThm}}
Proof of Theorem~\ref{thm:mainThm}: 
The encoding and decoding functions for each time-slot is same as given in Section~\ref{sec:codeD}. The primary state process $\{S_t\}$ is assumed to be an irreducible, aperiodic, finite-state homogeneous Markov chain and is independent of the secondary channel's input and outputs. 

Achievability follows from Theorem~$1$ in \cite{vis99}, when the state is perfectly known. We give an outline here.

\subsubsection{Achievability}
During the $i^{\text{th}}$ block, with the starting state $s_0$, the encoder chooses the $i^{\text{th}}$ symbol of the $T$ codewords of length $\pi(s_0)B$ for $T$ messages and multiplexes them $\bm{\rho}_i(s_0)=(\rho_1(s_0),\ldots,\rho_T(s_0)), i = 1 \ldots, B$. For large $B$ the capacity of the component channels $C_{s_0}(t)=\max\limits_{p(x|s_0)} \sum\limits_{s} p(s|s_{0}) I(X;Y|s, s_0)$ can be achieved. The final rate is calculated by summing over the $T$ multiplexed codewords, $C=\sum\limits_{s_0}\frac{\pi(s_0)}{T+1}\sum\limits_{t=1}^{T}C_{s_0}(t)$ which concludes the achievability.


\subsubsection{Converse}
Let $W$ be the message random variable. The capacity can be written as
\begin{equation*}
(T+1)C=\limsup_n \frac{1}{n} \sum\limits_{s_0}\pi(s_0)\sum_{t=1}^{T} \log M \stackrel{(a)}{\leq} \limsup_n \frac{1}{n} \sum\limits_{s_0}\pi(s_0)  \max_{\rho_1^{(t)n}} \sum_{t=1}^{T} I(W^{(t)};Y_1^{(t)n},S_0,S_1^{(t)n}),
\end{equation*}
where $(a)$ follows the same steps as in Equation (4) of \cite{vis99}, and the superscript $(t)n$ refers to time slot $t$ of block $n$. Now, 
\begin{align*}
\sum_{t=1}^{T} I(W^{(t)};Y_1^{(t)n},S_0,S_1^{(t)n})
\stackrel{(b)}{= }{} & \sum_{t=1}^{T} I(W^{(t)};Y_1^{(t)n}|S_0,S_1^{(t)n})\\
= {} & \sum_{t=1}^{T} \sum_{i=1}^{n} \left[ H(Y_i^{(t)}|S_0^{(i)},S_1^{(t)n}, Y_1^{(t)i-1}) - H(Y_i^{(t)}|S_0^{(i)},S_1^{(t)n}, W^{(t)}, Y_1^{(t)i-1}) \right]\\
\stackrel{(c)}{\leq} {} & \sum_{t=1}^{T} \sum_{i=1}^{n} \left[ H(Y_i^{(t)}|S_0^{(i)},S_i^{(t)}) - H(Y_i^{(t)}|S_0^{(i)},S_i^{(t)},X_i^{(t)}) \right]\\
= {} & \sum_{t=1}^{T} \sum_{i=1}^{n} I (X_i^{(t)};Y_i^{(t)} | S_0^{(i)}, S_i^{(t)}),
\end{align*}
where $(b)$ follows from the fact that the state is independent of the message and $(c)$ follows from the fact that entropy decreases on conditioning and $Y_i$ is independent of other random variables when conditioned on $W_i $ and $S_0^{(i)}$. Finally,
\begin{align*}
C 
\leq {} & \limsup_n 
\sum\limits_{s_0}\frac{\pi(s_0)}{n(T+1)} \sum_{t=1}^{T} \max_{X_1^{(t)n}} \sum_{i=0}^{n-1}  I(X_i^{(t)};Y_i^{(t)}| S_0^{(i)}, S_i^{(t)}) \\
\stackrel{(d)}{=} {} & \limsup_n 
\sum\limits_{s_0}\frac{\pi(s_0)n}{n(T+1)}  \sum_{t=1}^{T} \max_{p(X^{(t)}|S_0)}  I(X^{(t)};Y^{(t)}| S_0, S^{(t)}) \\
\leq {} & 
\sum\limits_{s_0}\frac{\pi(s_{0})}{T+1} \sum_{t=1}^{T}  \max_{p(X^{(t)}|S_0)}I(X^{(t)};Y^{(t)}| S_0,S^{(t)}),
\end{align*}
where $(d)$ follows the same steps as Equation (8) in \cite{vis99} which completes the proof. 


\section{\label{proof:PSol}}
Proof of Theorem~\ref{thm:PSol}:
The solution can be rewritten as,
\begin{equation}
\rho_{sSt}^* = \left\{ \begin{array}{ll} 
0 & \mbox{if $\text{SIC}_\text{C} < 0$} \\ 
\text{SIC}_\text{C} & \mbox{if $0 \leq \text{SIC}_\text{C} < \min\left( \text{SNR}_2, \text{INR}_\text{C} \right)$} \\
\min\left( \text{SNR}_2, \text{INR}_\text{C} \right) & \mbox{if $\text{SIC}_\text{C} \geq \max\left( \text{SNR}_2, \text{INR}_\text{C} \right)$}
\end{array}.
\right.
\end{equation}

The first part follows trivially from the definition of the problem. To prove the other parts, let the optimal solution to the optimization problem be given by $(\bm{\rho}_0^*, \bm{\rho}_1^*)$ such that $\rho_{sSt}^* < \text{SIC}_\text{C}$ for some $t\in [1,T]$ and $s\in\{0,1\}$. Let us consider $\rho_{sNt}^*, \rho_{sSt}^*$, keeping all other powers fixed. The constraints for this pair (except the positivity and complementarity constraints) are
$A+\pi(s)(\rho_{sNt}^* + \rho_{sSt}^*) \leq T\text{SNR}_2, 
\rho_{sSt}^*  \leq \text{SIC}_\text{C}, \text{and }
\rho_{sNt}^* + \rho_{sSt}^*  \leq \text{INR}_\text{C}$
, where the contribution to the average power from all other powers except $(\rho_{sNt}^*, \rho_{sSt}^*)$ are lumped into $A$. The rate can be written as,
\begin{equation*}
R_2(\rho_{sNt}^*, \rho_{sSt}^*) = R_2^{(0)}+ \pi(s)\left[\bar{\beta}_s(t)\mathcal{C}(\rho_{sNt}^* + \rho_{sSt}^*) + \beta_s(t)\left( \mathcal{C}\left( \frac{\rho_{sNt}^*}{1+\text{INR}_2+\rho_{sSt}^*} \right) + \mathcal{C}(\rho_{sSt}^*)\right)\right].
\end{equation*}
The contribution due to all other powers to the average rate has been lumped into $R_2^{(0)}$. Now let us take away a small part of the power from $\rho_{sNt}^*$ and add it to $\rho_{sSt}^*$. This has no effect on the sum $\rho_{sNt}^* + \rho_{sSt}^*$, hence the constraints remain unchanged. We chose a $\delta>0$ such that $\rho_{sNt}^*-\delta\geq 0$ and $\rho_{sSt}^*+\delta \leq \text{SIC}_\text{C}$. 
It is easy to see that $R_2(\rho_{sNt}^*, \rho_{sSt}^*) < R_2(\rho_{sNt}^*-\delta, \rho_{sSt}^*+\delta)$,
\begin{align*}
 \text{i.e. }\mathcal{C}\left( \frac{\rho_{sNt}^*}{1+\text{INR}_2+\rho_{sSt}^*} \right) + \mathcal{C}(\rho_{sSt}^*) &< \mathcal{C}\left( \frac{\rho_{sNt}^*-\delta}{1+\text{INR}_2+\rho_{sSt}^*+\delta} \right) + \mathcal{C}(\rho_{sSt}^*+\delta)\\
 \text{i.e. }\left( 1+\frac{\rho_{sNt}^*}{1+\text{INR}_2+\rho_{sSt}^*} \right)(1+\rho_{sSt}^*) &< \left( 1+\frac{\rho_{sNt}^*-\delta}{1+\text{INR}_2+\rho_{sSt}^*+\delta} \right)(1+\rho_{sSt}^*+\delta)\\
 \text{i.e. }\frac{(1+\text{INR}_2+\rho_{sNt}^*+\rho_{sSt}^*)(1+\rho_{sSt}^*)}{1+\text{INR}_2+\rho_{sSt}^*} & < \frac{(1+\text{INR}_2+\rho_{sNt}^*+\rho_{sSt}^*)(1+\rho_{sSt}^*+\delta)}{1+\text{INR}_2+\rho_{sSt}^*+\delta}
\end{align*}
since $\text{INR}_2>0$. 
This perturbation can only go on till either $\rho_{sNt}$ reaches its lower limit, 
or $\rho_{sSt}$ reaches the upper limit.
We hit the lower limit of $\rho_{sNt}$ first if $\text{SIC}_\text{C} \geq \max \left( \text{SNR}_2, \text{INR}_\text{C}\right)$. In this case, $\rho_{sNt}^*=0$ and $\rho_{sSt}=\min\left( \text{SNR}_2, \text{INR}_\text{C} \right).$
Otherwise, we hit the upper limit $\rho_{sSt}$ first, if $\rho_{sSt}=\text{SIC}_\text{C}$. 
In either case, the result is independent of the particular value of $t$.

\section{\label{proof:nonB}}
Proof of Theorem~\ref{thm:nonBCap}: 
The only variable is the superposition fraction $\alpha$, so the problem is one dimensional. The derivative of the optimization function is given by,
$\frac{dR_2(\alpha)}{d\alpha}
=\frac{\text{INR}_2\text{SNR}_1\text{SNR}_2}{(1+\alpha \text{SNR}_2)(1+\text{INR}_2+\alpha \text{SNR}_2)}
$.
The slope of the optimization function is always positive, i.e. $\frac{dR_2(\alpha)}{d\alpha} > 0$. Hence, the optimal solution is achieved at the upper boundary of the intersection of the constraints on $\alpha$ which are $0 \leq \alpha \leq 1$ and $\alpha \leq \frac{\frac{\text{INR}_2}{\text{SNR}_1}(1+\text{INR}_\text{gap})-1}{\text{SNR}_2}$. If the right hand side is less than or equal to $0$, then $\alpha^*=0$, if the right hand side is more than $1$, $\alpha^*=1$, otherwise, $\alpha^*=\frac{\frac{\text{INR}_2}{\text{SNR}_1}(1+\text{INR}_\text{gap})-1}{\text{SNR}_2}$. For $|h_{12}|=1$, this is same as the rate splitting result for two users as shown in \cite{ru96}.
\paragraph{If $\text{INR}_2 \geq \frac{\text{SNR}_1(1+\text{SNR}_2)}{1+\text{INR}_\text{gap}}$}
This comes from the decodability condition
$ \mathcal{C}\left( \frac{\text{INR}_2}{1+\text{SNR}_2} \right) \geq \mathcal{C}\left( \frac{\text{SNR}_1}{1+\text{INR}_\text{gap}} \right).$
Hence, after decoding off the primary, the rate that can be achieved by the secondary is $R_{2}(\alpha^{*})=\mathcal{C}\left( \text{SNR}_2 \right)$. But this is the single user capacity of the secondary channel for the given power constraint when there is no primary interference. Therefore, this scheme of using only a Layer~$2$ codeword is optimal in this case.

\paragraph{If $\text{INR}_2 \leq \frac{\text{SNR}_1}{1+\text{INR}_\text{gap}}$} the primary cannot be decoded even if no secondary data is sent i.e. $ \mathcal{C}\left( \text{SNR}_2 \right)   \leq \mathcal{C}\left( \frac{\text{SNR}_1}{1+\text{INR}_\text{gap}} \right)$. So, in this case, the only option is to treat this undecodable signal as noise. The effective channel now behaves as a point to point channel with AWGN $\mathcal{N}(0,1+\text{INR}_2)$. Hence the capacity of this channel is given by $R_{2}(\alpha^{*})=\mathcal{C}\left( \frac{\text{SNR}_2}{1+\text{INR}_2} \right)$.
\paragraph{If $\frac{\text{SNR}_1(1+\text{SNR}_2)}{1+\text{INR}_\text{gap}} < \text{INR}_2 < \frac{\text{SNR}_1}{1+\text{INR}_\text{gap}}$} For the rest of the parameter range, the rate is a corner point of the 3 user virtual MAC formed by the primary and the two layers of the secondary (Figure~\ref{fig:N1geqN2A}). The fixed primary transceiver converts the effective MAC (from the point of view of the secondary receiver) into a straight line and the secondary splits its information into two layers to achieve the boundary point. The sum capacity of the effective MAC is achieved by this scheme since for any $\alpha$, $\mathcal{C}\left( \frac{(1-\alpha)\text{SNR}_{2}}{1+ \text{INR}_{2}+\alpha \text{SRN}_{2}} \right) + \mathcal{C}\left( \frac{\text{INR}_{2}}{1+\alpha \text{SNR}_{2}} \right) + \mathcal{C}\left( \alpha \text{SNR}_{2}\right) = \mathcal{C}\left( \text{INR}_{2} + \text{SNR}_{2}\right)$. Additionally $\alpha^{*}$ gives the best rate among all the secondary rates. Hence the layered encoding at the secondary transmitter and sequential decoding at the secondary receiver is optimal and the capacity is given by $R_{2}(\alpha^{*})$. 

\bibliographystyle{IEEEtranS}
\bibliography{references_dash}
\end{document}